 \documentclass[apj,numberedappendix]{emulateapj}
  \usepackage{amsmath}

\def\simlt{\lower.5ex\hbox{$\; \buildrel < \over \sim \;$}}
\def\simgt{\lower.5ex\hbox{$\; \buildrel > \over \sim \;$}}

\def\beq{\begin{equation}}
\def\eeq{\end{equation}}
\def\ba{\begin{eqnarray}}
\def\ea{\end{eqnarray}}
\def\bB{\boldsymbol{B}}
\def\bE{\boldsymbol{E}}

\def\bv{\boldsymbol{v}}
\def\bOm{\boldsymbol{\Omega}}
\def\bk{\boldsymbol{k}}

\def\Sect{{\rm Section}} 
\def\Sects{{\rm Sections}} 

\def\Eq{Equation}
\def\Eqs{Equations}

\def\M{{\cal M}}
\def\Lsd{L_{\rm w}}
\def\sT{\sigma_{\rm T}}

\def\IGJ{I_0}

\def\tauT{\tau_{\rm T}}
\def\texp{t_{\rm exp}}

\def\B{{\cal B}}

\def\Gf{\Gamma_{\rm f}}
\def\Lf{L_{\rm f}}
\def\Gw{\Gamma_{\rm w}}
\def\Lw{L_{\rm w}}
\def\Bw{B_{\rm w}}

\def\sigw{\sigma_{\rm w}}

\def\rL{r_{\rm L}}
\def\nuobs{\nu_{\rm obs}}

\def\eff{\epsilon}

\def\E{{\cal E}}
\def\N{{\cal N}}
\def\I{{\cal I}}
 \def\Rdec{R_\diamond}
\def\nudec{\nu_\diamond}
\def\tdec{t_\diamond}
\def\Gdec{\Gamma_\diamond}
\def\Ldec{L_\diamond}
\def\Edec{\E_\diamond}
\def\adec{a_\diamond}

\def\dbw{\Delta_{\rm bw}}

\def\Grel{\Gamma_{\rm rel}}

\def\RLC{R_{\rm LC}}
\def\dN{\dot{N}}

\def\Ebw{{\cal E}_{\rm bw}}

\def\tobs{t_{\rm obs}}

\def\EFRB{{\cal E}_{\rm FRB}}
\def\Bf{B_{\rm f}}
\def\trep{t}

\def\tc{t_c}
\def\nucom{\tilde{\nu}}
\def\nupeak{\nu_{\rm peak}}

\def\omL{\omega_{\rm L}}

\def\ThB{\Theta_{B}}

\def\nupk{\nu_{\rm peak}}

\def\tth{\tilde{\theta}}

\def\tBw{\tilde{B}_{\rm w}}

\def\Eb{{\cal E}_{\rm b}}
\def\RFS{{\cal R}_{\rm F}}
\def\RRS{{\cal R}}
\def\vFS{v_{\rm F}}
\def\Et{{\cal E}_{\rm t}}
\def\gth{\gamma_{\rm th}}
\def\gb{\gamma_{\rm b}}

\def\RA{R_{\rm A}}
\def\BA{B_{\rm A}}
\def\Gsh{\Gamma_{\rm sh}}

\def\Icl{I}
\def\dNcl{\dN_{\rm tw}}

\def\taugg{\tau_{\gamma\gamma}}
\def\ti{t_i}
\def\LFRB{L_{\rm FRB}}
\def\thb{\theta_b}
\def\tthb{\tilde{\theta}_b}
\def\Omb{\Omega_b}
\def\tOmb{\tilde{\Omega}_b}
\def\tThB{\tilde{\Theta}_B}

\def\bOm{\boldsymbol{\Omega}}

\def\vsh{v_{\rm sh}}

\def\RNS{R_{\rm NS}}
\def\BNS{B_{\rm NS}}

\newbox\grsign \setbox\grsign=\hbox{$>$} \newdimen\grdimen \grdimen=\ht\grsign
\newbox\simlessbox \newbox\simgreatbox \newbox\simpropbox
\setbox\simgreatbox=\hbox{\raise.5ex\hbox{$>$}\llap
     {\lower.5ex\hbox{$\sim$}}}\ht1=\grdimen\dp1=0pt
\setbox\simlessbox=\hbox{\raise.5ex\hbox{$<$}\llap
     {\lower.5ex\hbox{$\sim$}}}\ht2=\grdimen\dp2=0pt
\setbox\simpropbox=\hbox{\raise.5ex\hbox{$\propto$}\llap
     {\lower.5ex\hbox{$\sim$}}}\ht2=\grdimen\dp2=0pt
\def\simgt{\mathrel{\copy\simgreatbox}}
\def\simlt{\mathrel{\copy\simlessbox}}

\begin{document}

\title{Blast waves from magnetar flares and fast radio bursts}

\author{
Andrei M. Beloborodov
}
\affil{$^1$Physics Department and Columbia Astrophysics Laboratory,
Columbia University, 538  West 120th Street New York, NY 10027\\
$^2$Max Planck Institute for Astrophysics, Karl-Schwarzschild-Str. 1, D-85741, Garching, Germany
% amb@phys.columbia.edu
}

\begin{abstract}
Magnetars younger than one century are expected to be hyperactive: besides winds powered by rotation they can generate frequent magnetic flares, which launch powerful blast waves into the wind. The resulting magnetized shocks act as masers producing bright radio emission. This theoretical picture predicts radio bursts with the following properties. (1) GHz radio emission occurs at radii $r\sim 10^{14}$~cm and lasts $\simlt 1$~ms in observer's time. (2) Induced scattering in the surrounding wind does not suppress the radio burst. (3) The emission has  linear polarization set by the magnetar rotation axis. (4) The emission drifts to lower frequencies during the burst, and its duration broadens at lower frequencies. (5) Blast waves in inhomogeneous winds may emit variable radio bursts; periodicity might appear on sub-ms timescales if the magnetar rotates with $\sim 1$~s period. However, the observed burst structure is likely changed by lensing effects during propagation through the host galaxy. (6) The magnetar bursts should repeat, with rare ultrastrong events (possibly up to $\sim 10^{43}$~erg in radio waves) or more frequent weak bursts. (7) When a repeating magnetic flare strikes the wind bubble in the tail of a previous flare, the radio burst turns into a bright optical burst lasting $\simlt 1$~s. Locations of hyper-active magnetars in their host galaxies depend on how they form: magnetars created in supernovae explosions will trace star formation regions, and magnetars formed in mergers of compact objects will be offset. The merger magnetars are expected to be particularly hyper-active.
\end{abstract}


%#####################################################################

\section{Introduction}

Hyper-active magnetars (HAMs) are hypothetical objects that produce giant 
flares much more frequently than the older local magnetars detected in our Galaxy. The hyper-activity is expected during accelerated ambipolar diffusion in the core of a young magnetar, and is hard to catch because this evolution phase is short. Recent estimates suggest that magnetic energy release peaks during the first $\sim 10^9$~s of a magnetar life \citep{Beloborodov16}, which should lead to enormous flaring activity. All of the $\sim 30$ observed magnetars are much older (their typical age is $\sim 10^{11}$~s), and their activity is relatively modest --- so far only 3 giant flares have been detected from the entire population of local magnetars.

Besides a young age, two factors can accelerate ambipolar diffusion (\citealt{Beloborodov17b}, hereafter Paper~I). (1) Sufficiently massive magnetars cool faster due to activation of direct Urca reactions, and ambipolar diffusion accelerates in response to increased neutrino cooling. (2) Extremely strong magnetic fields make ambipolar diffusion particularly fast (e.g. \cite{Goldreich92}). The strongest fields are expected in neutron stars born with unusually high spins (ms periods), as the field can be strongly amplified by differential rotation \citep{Spruit08} and dynamo \citep{Duncan92}. The ultra-fast rotation is most likely in magnetars formed in mergers. It is also possible in magnetars formed by collapsing massive stars of low metallicity, because  such progenitors have weaker winds and retain higher angular momenta.

While still hypothetical, HAM is a plausible candidate for the engine of the 
repeating FRB~121102 (Paper~I), and a similar scenario might explain also other fast radio bursts (FRBs, see \cite{Petroff19} for a review). In the model proposed in Paper~I, the radio bursts come from ultra-relativistic blast waves launched into the magnetar wind by repeating magnetic flares.

Paper~I also proposed that the persistent radio nebula associated with FRB~121102 was inflated by ion ejecta from the magnetar flares. Both energy and particle content of the nebula were found consistent with this scenario, calibrated by observations of ejecta from SGR~1806-20. The electron-ion nebula can affect the polarization vector of the bursts through Faraday rotation. \cite{Margalit18} proposed that this may explain the large rotation measure ${\rm RM}\sim 10^5$~rad/m$^2$ observed in FRB~121102 \citep{Michilli18}. The possible polarization effect of the nebula may be further clarified by  detailed (multi-zone) transfer models, including partial conversion to circular polarization \citep{Vedantham19,Gruzinov19}.

Paper~I proposed that the blast wave in the magnetar wind emits a radio burst by the well-known maser mechanism. Maser emission from a relativistic magnetized shock 
was previously discussed in the context of the Crab nebula and applied to the pulsar wind termination shock \citep{Hoshino91,Gallant92}. \cite{Lyubarsky14} considered a similar termination shock for a magnetar wind and pointed out that a giant flare could greatly boost the power of the termination shock and produce an FRB (see also \cite{Murase16}). However, this scenario was found to be in tension with observations of FRB~121102, because it required an unrealistic energy budget for the frequent repeater, and because the long recovery time of the termination shock would prevent frequent bursts. Instead, Paper~I suggested a train of multiple internal shocks in the magnetar wind, which result from multiple magnetospheric flares. These shocks emit FRBs at radii $10^3-10^4$ times smaller than the radius of the wind termination shock. Paper~I also pointed out that in frequent repeaters the blast wave can strike the slow tail of ion matter ejected in a previous flare. \cite{Metzger19} recently developed a detailed model for blast waves in such tails and suggested that their emission reproduces the observations of FRB~121102. Their conclusions, however, differ from the results presented below (the comparison is given in \Sect~8.2).

The present paper systematically investigates the expected properties of FRBs from magnetar blast waves. It first describes magnetar winds and then blast waves from giant flares, and how they can produce radio and optical bursts. We do not discuss or compare here numerous other scenarios proposed for the FRB phenomenon. They are quite diverse, ranging from coherent emission of charge bunches inside a neutron star magnetosphere (e.g. \cite{Katz16,Kumar17}) to neutron stars interaction with an external plasma flow \citep{Zhang18} to collisions of microscopic magnetic dipoles \citep{Thompson17}. 

The paper is organized as follows. Section~2 describes persistent, rotationally driven winds from young magnetars, and Section~3 examines the wind structure in the presence of intermittent ejection of heavy ion material. The wind parameters are important for our FRB model, because they determine both dynamics and luminosity of the blast waves in the wind. The blast wave is driven by a magnetic plasmoid ejected by a magnetospheric giant flare, as described in \Sect~4. \Sect~5 estimates synchrotron radiation from the blast wave and examines conditions for a bright optical flash. Coherent radio emission from the blast wave is discussed in detail in Section~6, including the strength parameter of the radio wave, its beaming, polarization, time profile of the radio burst, and its spectral evolution. Constraints imposed by induced Compton scattering are described in \Sect~7. Our conclusions are summarized in \Sect~8.

%###################################################################

 \section{Rotationally driven wind}
\label{wind}

\subsection{Rotation rate and spindown power}
\label{sd}

Persistent winds from magnetars are powered by their rotation, similar to ordinary pulsar winds. 
A neutron star with angular velocity $\bOm$ and magnetic dipole moment $\boldsymbol{\mu}$ 
generates the net Poynting flux \citep{Goldreich69},
\beq
\label{eq:sd}
  \Lsd\approx \frac{\mu^2\Omega^4}{c^3}\approx 10^{37}\,\mu_{33}^2
    \left(\frac{\Omega}{4\,{\rm rad/s}}\right)^4 \,\frac{\rm erg}{\rm s}.
\eeq
Here we normalized $\mu$ to $10^{33}\,$G$\,$cm$^3$, typical for magnetars in our galaxy. It corresponds to surface magnetic field $\BNS\sim \mu/\RNS^3\sim 10^{15}\mu_{33}\,$G, where $\RNS\sim 10^6$~cm is the neutron star radius.\footnote{The relation $\BNS\sim \mu/\RNS^3$ assumes a dipole magnetic field on the surface. The actual surface field 
is likely non-dipolar and can be much stronger. The field inside the neutron star is yet stronger.}
We also normalized $\Omega$ to 4~rad/s, because magnetars with ages of interest in this paper are expected to slow down to $\Omega\sim 2-10$~rad/s. They lose rotation energy with rate $\I\Omega \dot{\Omega}=-\Lsd$, where $\I\approx 10^{45}$~g~cm$^2$ is the neutron star's moment of inertia. The simplest model where $\mu$ is taken constant in time (which may be a rough approximation for active magnetars), or is replaced by its time-averaged value $\bar{\mu}$, gives the following estimate for the rotation period at age $t$,
\beq
  P=\frac{2\pi}{\Omega}\approx 2\pi\,\bar{\mu}\left(\frac{2\,t}{c^3\I}\right)^{1/2}
   \approx 1.7\,\bar{\mu}_{33}\,t_9^{1/2} {\rm~s}.
\eeq 

Copious $e^\pm$ creation in the magnetosphere (estimated in \Sect~\ref{pairs} below) implies that the spindown energy flow from the star, $\Lsd$, may be approximately described as an MHD wind. The $e^\pm$ plasma tends to screen electric fields $E_\parallel$ parallel to $\bB$, leading to nearly ``force-free'' dynamics of the magnetosphere co-rotating with the star. This allows the field lines to open at the light cylinder,
\beq
   \RLC=\frac{c}{\Omega},
\eeq
and form a magnetized plasma wind that was previously studied in detail in the context of ordinary radio pulsars.
The magnetic field lines are bent around the rotation axis, so that the toroidal field $B_\phi$ at $\RLC$ is comparable to the poloidal field $B_{\rm pol}\sim\mu/\RLC^3$ \citep{Goldreich69}.
At radii $r\gg \RLC$ the wind becomes nearly radial, and $B_{\rm pol}\approx B_r$ becomes much smaller than the toroidal magnetic field, $|B_\phi/B_r|\approx r/\RLC$ (see \cite{Kirk09,Cerutti17} for reviews).

\subsection{Electron-positron loading}
\label{pairs}

The toroidal field at the light cylinder $B_\phi\sim \mu/\RLC^3$ requires electric current along the open magnetic field lines,
\beq
  \IGJ\approx \frac{c\mu}{\RLC^2} \approx \left(c\Lsd\right)^{1/2}.
\eeq
In ordinary pulsars, the rate of particle outflow in the wind is usually described as a multiple of $\IGJ/e$,
\beq
  \dN_0=
   \M\,\frac{\IGJ}{e} \approx 10^{37}\,\M_3\, L_{\rm w,37}^{1/2} {\rm ~s}^{-1}.
\eeq
Here $\M$ is a multiplicity coefficient due to copious pair creation on the open 
magnetic field lines. Kinetic plasma simulations of $e^\pm$ discharge in pulsars show that $\dN_0$ is concentrated around the  current sheet separating the closed and open field lines and extending beyond the light cylinder \citep{Chen14,Philippov15}. Most of $e^\pm$ pairs ejected by young pulsars are created through photon-photon collisions at  $r\sim\RLC$. 

Magnetars are significantly different. During their hyper-active youth, magnetars have a persistently twisted {\it closed} magnetosphere, which carries electric current $\Icl\gg \IGJ$. Therefore, the rate of pair creation by magnetars greatly exceeds that from $\IGJ$. Most of the pairs are created on closed field lines near the star, however a fraction estimated below end up in the open field-line bundle and form a plasma wind flowing out through the light cylinder.

The closed twist current $I$ is sustained through $e^\pm$ discharge \citep{Beloborodov07,Beloborodov13a}. Pairs are created by photons (originally emitted by the star) that are resonantly upscattered in the magnetosphere and convert to $e^\pm$. The photon conversion is nearly instantaneous in the inner magnetosphere where $B>10^{13}$~G, resulting in an $e^\pm$ avalanche. The avalanche develops at radii up to $\sim R_\pm$ where 
\beq
\label{eq:Rpm}
    B(R_\pm)=10^{13}\,{\rm G}=B_\pm.
\eeq
The net current along twisted closed field lines extending to $r\simgt R_\pm$ is  
\beq
   \Icl\sim \psi\,c R_\pm B_\pm,
\eeq
where $\psi$ is the twist angle, which is comparable to unity for active magnetars. 

To get a numerical estimate for the current $I$, consider a twisted dipole magnetosphere. Then $R_\pm \sim 5\times 10^6 \mu_{33}^{1/3} {\rm~cm}$, and $I(R_\pm)\sim \psi (\RLC/R_\pm)^2 I_0$ exceeds $I_0$ by a factor of $\sim 10^6$. The actual current may be stronger if the magnetosphere is far from dipole and has ultra-strong flux tubes emerging from the surface, like sunspots.

\begin{figure}[t]
\begin{center}
\includegraphics[width=0.43\textwidth]{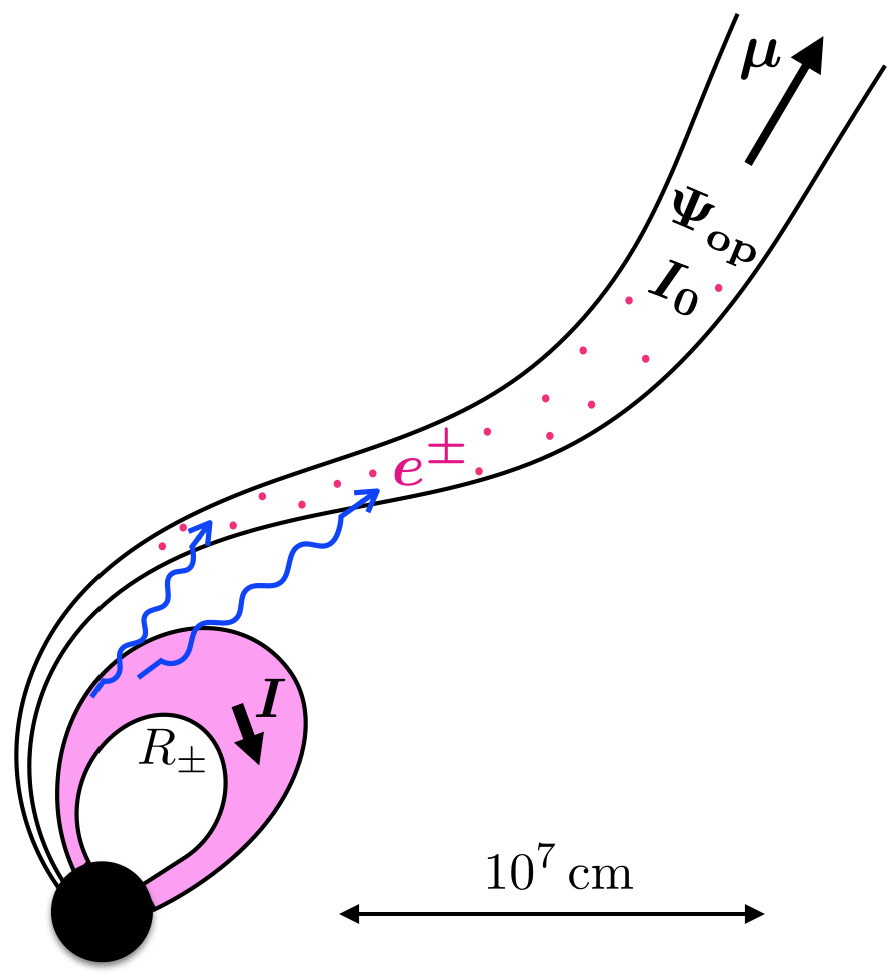} 
\end{center}
\caption{Mechanism of $e^\pm$ loading of the open field-line bundle. The  avalanche of most numerous $e^\pm$ creation develops in the closed magnetic loops extending to $r\sim R_\pm\sim 50$~km (\Eq~\ref{eq:Rpm}) and carrying current $I\sim 10^6 I_0$ (shaded in magenta). The loop sprays copious gamma-rays (blue arrows), and some of them convert to $e^\pm$ in the open bundle (magenta dots). At large radii, the open bundle (magnetic flux $\Psi_{\rm op}$) becomes axisymmetric about the direction of the magnetic dipole moment $\boldsymbol{\mu}$. However, its shape in the inner magnetosphere $r<10^7\,$cm can be complicated, with multiple legs on the stellar surface.
}
\label{fig:pair_loading}
 \end{figure}

Twisted magnetic loops that extend beyond $R_\pm$ produce $e^\pm$ pairs with highest multiplicity, as the number of created pairs is dominated by the last generation of the $e^\pm$ avalanche with Lorentz factors $\gamma(R_\pm)\sim 10$ \citep{Beloborodov13a}. The last generation is produced by photons with $l_{\rm ph}\sim R_\pm$, which spray $e^\pm$ throughout the magnetosphere.
The rate of pair creation in this spray is 
\beq
\label{eq:dNtw}
   \dNcl\sim \M\,  \frac{\Icl}{e}\sim \M \,\frac{c\, R_\pm B_\pm}{e}.
\eeq
The multiplicity factor $\M$ depends on the voltage $\Phi$ that sustains the twist current,
\beq
   \M\sim \frac{e\Phi}{\gamma(R_\pm) m_ec^2}\sim 10^3\,\left(\frac{e\Phi}{10\,\rm GeV}\right).
\eeq
A fraction of the spayed $e^\pm$ will end up in the open field line bundle (Figure~\ref{fig:pair_loading}). This fraction scales with the volume $V$ occupied by the open bundle at radii $r\sim R_\pm$,
\beq
\label{eq:frac}
     \frac{\dN}{\dNcl}=\zeta\,\frac{V}{2\pi R_\pm^3},
\eeq
where $\zeta$ is a geometric coefficient determined by the shape of the open bundle and its position relative to the twisted closed loops. In a weakly deformed dipole magnetosphere $\zeta$ would be small (it can be explicitly calculated taking into account that the gamma-rays are emitted tangentially to the magnetic field lines, see \cite{Beloborodov13a}). However, realistic HAM magnetospheres have curved open bundles, possibly with multiple legs. Then $\zeta$ may exceed unity. 

%%%%%%%%%%% FIGURE %%%%%%%%%%%%%%%%%%
%%%%%%%%%%% FIGURE %%%%%%%%%%%%%%%%%%

The open bundle volume at $r\sim R_\pm$ is related to its magnetic flux $\Psi_{\rm op}$,
\beq
\label{eq:Vop}
   V\sim R_\pm\,\frac{\Psi_{\rm op}}{B(R_\pm)}, \qquad 
   \Psi_{\rm op}\approx \frac{2\pi \mu}{\RLC},
\eeq
where $B(R_\pm)$ is comparable to $B_\pm=10^{13}$~G.\footnote{At radii $r\sim 10^7\,{\rm cm}\gg R_\star$, $B$ in the open bundle is not vastly different from $B$ in the closed loops. Even a strongly intermittent  magnetosphere (with localized ``sunspots'') at large $r$ approaches a quasi-dipole configuration with a smooth distribution of $B$.}
\Eqs~(\ref{eq:dNtw}), (\ref{eq:frac}), and (\ref{eq:Vop}) give the following order-of-magnitude estimate for the rate of $e^\pm$ outflow through the light cylinder,
\beq
\label{eq:dN}
  \dot{N}\sim \zeta \, \M\,\frac{\mu\Omega}{eR_\pm}
  \sim \frac{3\times 10^{39}}{P}\,\zeta \,\M_3\,\mu_{33}^{2/3}.
\eeq

Another source of $e^\pm$ pairs is the equatorial current sheet outside $\RLC$, where magnetic reconnection occurs. Particles accelerated by reconnection emit synchrotron and inverse Compton (IC) photons, and a fraction of the energetic IC gamma-rays convert to $e^\pm$ via photon-photon collisions; simulations of this process were recently performed by \cite{Hakobyan19}.

Let $\epsilon_{\rm diss} \Lw$ be the fraction of spindown power dissipated near $\RLC$ and converted to accelerated particles. Kinetic plasma simulations of pulsars suggest that $\epsilon_{\rm diss}$ can approach $\sim 0.2$ (see \cite{Cerutti17} for a review). Reconnection releases energy of the order of $U_B/n$ per particle, where $U_B=B_{\rm LC}^2/8\pi$ and $n$ is the particle density near the light cylinder. This implies heating to  a characteristic Lorentz factor  $\gamma_{\rm rec}\sim \Lw/\dot{N}m_ec^2$. A large fraction of the dissipated power converts to synchrotron radiation in the magnetic field $B_{\rm LC}\sim \mu/\RLC^3\sim 10^4\,\mu_{33} P^{-3}$~G, where $P=2\pi/\Omega$ is measured in seconds. The synchrotron cooling timescale is short,
\beq
   t_s\sim \frac{m_ec}{U_B\,\sT\gamma_{\rm rec}}, \qquad 
   \frac{ct_s}{\RLC}\sim 0.1\,\left(\frac{\gamma_{\rm rec}}{10^3}\right)^{-1} \frac{P^5}{\mu_{33}^2}. 
\eeq
The synchrotron spectrum from the reconnection layer extends over a broad range of photon energies around $\gamma_{\rm rec}^2\hbar eB_{\rm LC}/m_ec$, from optical to the X-ray band, and the total synchrotron luminosity of the layer is $L_s\sim \epsilon_{\rm diss}\Lw$. It can be comparable to the magnetar luminosity $L$. 

The resulting radiation energy density at the light cylinder $U\sim (L+L_s)/4\pi\RLC^2c$ is comparable to $U_B$. Therefore, the IC losses of particles accelerated by reconnection are significant, even after taking into account their reduction by the Klein-Nishina effects. The fraction of particle energies lost to IC emission may be roughly estimated as $f_{\rm IC}\sim 0.1$. The IC photons have the characteristic energy $E_{\rm IC}\sim\gamma_{\rm rec} m_ec^2$, which is in the gamma-ray band. 

The IC gamma-rays can collide with the X-rays  flowing from the magnetar and turn into $e^\pm$ pairs. In particular, gamma-rays with energies $E_{\rm IC}\sim 0.1-1$~GeV collide most efficiently with X-rays of energies $E_X\sim 1-10$~keV photons, which are near the threshold for pair creation, $(E_{\rm IC}E_X)^{1/2}\sim 1$~MeV.  The cross section for such collisions is $\sigma_{\gamma\gamma}\sim 0.1\sT$, and the corresponding optical depth seen by the gamma-rays is given by
\beq
  \taugg\sim \sigma_{\gamma\gamma}n_X\RLC\sim \frac{0.1\sT L}{4\pi c \RLC  E_X}  \sim 0.1\, L_{37}.
\eeq
Here we substituted $\RLC\sim 5\times 10^{9}$~cm that corresponds to rotation period $P=1$~s. The X-ray luminosity of a young, hyperactive magnetar is likely much higher than $L\sim 10^{35}$~erg/s typical for the local (kyr-old) magnetars; therefore we normalized $L$ to $10^{37}$~erg/s. The synchrotron radiation with luminosity $L_s$ generated at the light cylinder may provide a comparable $\taugg$.

These estimates suggest that a small fraction $\epsilon_{\rm diss} f_{\rm IC}\tau_{\gamma\gamma}\sim 10^{-3}$ of the magnetar spindown power $\Lw$ converts to energetic $e^\pm$ around the light cylinder. The fraction $f_{\rm IC}\sim 0.1$ of the dissipated power $\epsilon_{\rm diss}\Lw$ converts to GeV photons, and the fraction $\taugg$ of the GeV photons convert to $e^\pm$ pairs with Lorentz factors $\sim 10^3$. These pairs are smoothly distributed throughout the wind. They cool to $\gamma_c\simlt 10^2$ (at which $t_s\sim \RLC/c$) and flow out in the wind. The particle number flux of this reconnection-powered outflow may be expressed in the form,
\beq
\label{eq:dNrec}
   \dN_{\rm rec} \sim \frac{\taugg\,f_{\rm IC}\,\epsilon_{\rm diss}\Lw}{E_{\rm IC}}
   \sim 10^{-2}\,  \frac{\taugg\,\Lw}{\gamma_{\rm rec}m_ec^2}.
\eeq 
It is likely smaller than $\dot{N}$ from the process shown in Figure~\ref{fig:pair_loading} and estimated in \Eq~(\ref{eq:dN}).

\subsection{Enhanced wind preceding a giant flare}

The blast wave from a magnetospheric explosion will interact with a small portion of the pre-explosion wind that was emitted seconds before the explosion (Paper~I). This pre-explosion wind can have power $\Lw$ and particle flux $\dot{N}$ much higher than the average spindown wind emitted between giant flares. The pre-flare wind enhancement is expected for two reasons.

(1) The effective $\boldsymbol{\mu}$ increases. Flares occur when the magnetosphere is over-twisted, $\psi\gg 1$. Before the flare (i.e. before magnetic reconnection and plasmoid expulsion) the magnetosphere inflates, and the open magnetic flux $\Psi_{\rm op}$ can far exceed its average value \citep{Parfrey13}. Then the effective $\mu\sim\Psi_{\rm op}/2\pi \RLC$ is temporarily increased, and $\dN$ grows $\propto \mu^{2/3}$ according to \Eq~(\ref{eq:dN}). 

(2)  The voltage $\Phi$ increases. The possibility of a high voltage in magnetars was discussed by \cite{Thompson08b}, who argued that plasma instabilities near the boundary between the twisted closed magnetosphere and the open field-line bundle increase $\Phi$. A strong increase of voltage cannot last long, because it leads to quick untwisting of the closed field lines \citep{Beloborodov09}. However $\Phi\gg 10$~GeV is possible for a short time preceeding a flare, if the pre-flare inflation of the magnetosphere triggers new instabilities and boosts the dissipation rate. Then $\dN$ grows according to \Eq~(\ref{eq:dN}), since $\M\propto\Phi$.

In numerical estimates in this paper we will normalize the pre-flare wind power $\Lw$ to $10^{39}\,$erg/s and the particle flow rate in the wind $\dN$ to $10^{42}\,$s$^{-1}$. We will keep track of how these poorly known parameters enter the final results. 

An important dimensionless parameter of the wind is its energy per unit rest mass,
\beq
\label{eq:eta}
  \eta\equiv \frac{\Lw}{\dot{N} m_ec^2}\sim 10^2-10^4.
\eeq
Here we assumed that the average mass per particle is close to electron mass $m_e$, neglecting ions that may flow from the ``polar caps'' (footprints of the open magnetic field lines on the star).
An upper limit on the pre-flare ion flow may be estimated as 
\beq
\label{eq:ions}
  \dot{N}_i\simlt \frac{\IGJ}{e}\sim 10^{35}\, L_{\rm w,39}^{1/2} {\rm ~s}^{-1}.
\eeq  
As the giant flare develops, a large number of ions can be lifted from the magnetar surface and ejected with a mildly relativistic speed, as suggested by observations of the December 2004 flare SGR~1806-20 \citep{Gaensler05}. However, this massive slow ejection occurs behind the ultra-relativistic blast wave from the flare, and normally does not contaminate the pre-flare wind. In frequent repeaters, a situation is possible where a flare blast wave runs into the slow tail of a preceding flare (Paper~I); this situation will be discussed in Section~\ref{impact}.

\subsection{Lorentz factor of the wind}

We will be mainly interested in the wind Lorentz factor $\Gw$ at large radii $r\gg \RLC$ (because it enters the blast wave model discussed in the subsequent sections). 
A simple estimate for $\Gw$ will be given in \Eq~(\ref{eq:Gw}).
This estimate uses the so-called ``force-free'' wind model.
It assumes that the wind emerges at the light cylinder with an initially sub-magnetosonic speed, and we need to check this assumption before using it.
Therefore, our first step is to evaluate the ejection speed of the wind plasma loaded in the open bundle at $r\ll \RLC$ (Figure~\ref{fig:pair_loading}). 
The key factor here is that the plasma is exposed to the radiation field of the magnetar and experiences a strong Compton drag before escaping the magnetosphere. This drag limits the outflow Lorentz factor $\gamma$. As a result, the wind emerges with a subsonic speed at the light cylinder. 

The radiation field around an active magnetar is shaped by radiative transfer due to resonant Compton scattering by magnetospheric particles. Detailed transfer simulations coupled to self-consistent $e^\pm$ dynamics were performed in \cite{Beloborodov13b}. They show that near the magnetic axis the resonant Compton drag decelerates the $e^\pm$ outflow to $\gamma\sim 2-3$ at radii $\sim 10R_\star$ (see Figure~9 in \cite{Beloborodov13b}). The star and its near magnetosphere (filled with scattered radiation and decelerated plasma) form an X-ray source of size $R_X\sim 2\times 10^{7}$~cm. 

At larger radii $r\gg R_X$ the X-rays decouple and flow out, approximately radially, within angle $\sim R_X/r\ll 1$. Here Compton drag tends to accelerate the $e^\pm$ outflow along the open field lines. However, the radiative force quickly decreases with radius, and $\gamma$ saturates before it could significantly grow. 
It is easy to verify that resonant Compton drag becomes inefficient at $r\sim 10^8$~cm, where $\hbar eB/m_ec\ll 1$~keV. Here resonant scattering becomes limited to low-energy photons in the Rayleigh-Jeans part of the magnetar spectrum, and these low-energy photons carry a negligible momentum flux, unable to apply a strong force to the plasma. The radiative force then comes only from non-resonant Thomson scattering of the main flux of photons with energies $E\sim 1-10 {\rm ~keV}\gg \hbar eB/m_ec$. 

In particular, consider the plasma near the magnetic dipole axis in the region $R_X\ll r\ll\RLC$. Here the $e^\pm$ outflow is almost radial, because the magnetic field lines are almost radial. Radiation flux $F\sim L/4\pi r^2$ is determined by  the magnetar luminosity $L$. An upper limit on the accelerating force $f$ is obtained assuming a perfectly collimated flux: $f=(\sT F/c)(1-\beta)/(1+\beta)$, which gives the acceleration timescale $t_{\rm acc}(\gamma)\sim 4m_ec^2\gamma^3/\sT F$ for $\gamma\gg 1$. Even for a moderate $\gamma\sim$ a few, one finds $ct_{\rm acc}/r\sim 16\pi m_ec^3\gamma^3 r/\sT L\gg 1$, and hence acceleration is inefficient.

We conclude that the main impact of Compton drag on the $e^\pm$ outflow occurs via resonant scattering at $r\sim R_X$ and sets the outflow Lorentz factor $\gamma\sim 2-3$. The outflow is cold, because Compton drag brings particles to approximately the same Lorentz factor, and because the subsequent expansion of the flow to $\RLC$ occurs with adiabatic cooling.

%%%%%%%%%%% FIGURE %%%%%%%%%%%%%%%%%%
\begin{figure*}[t]
% \vspace*{-8cm}
\begin{center}
\includegraphics[width=0.8\textwidth]{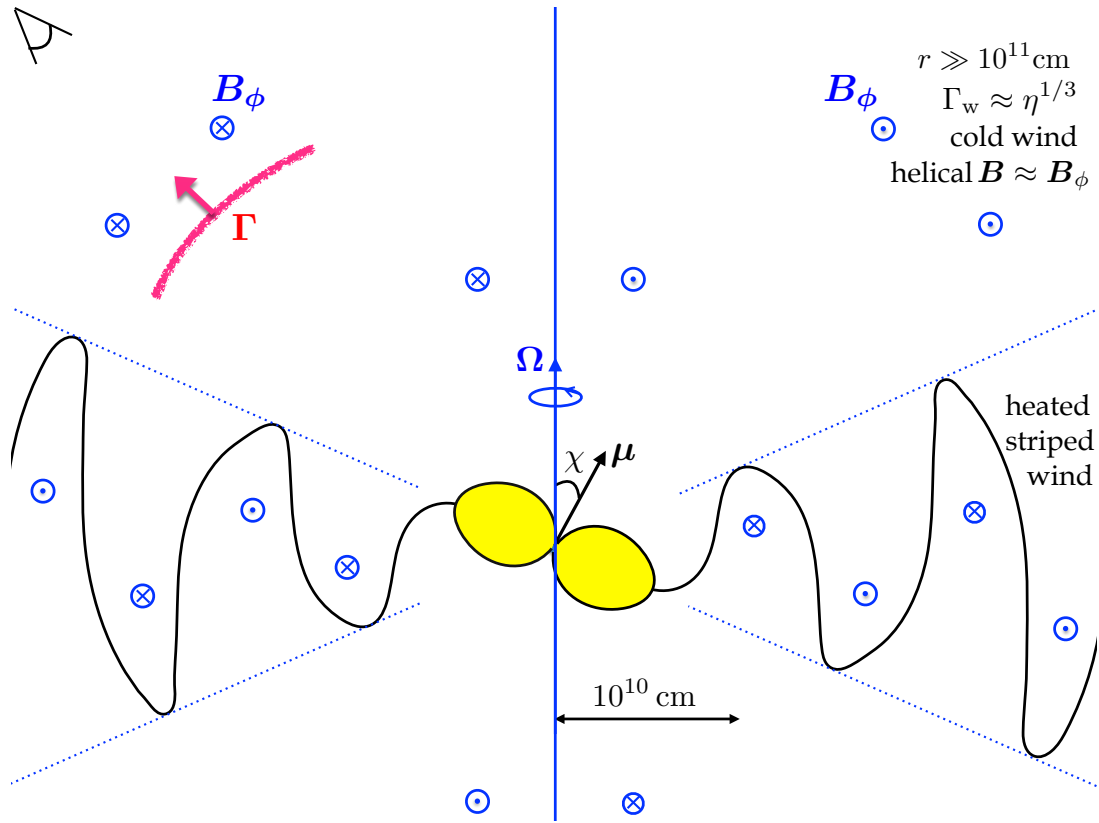} 
\end{center}
% \vspace*{-8.1cm}
\caption{Environment of an active magnetar with rotation period $P=2\pi/\Omega\sim 1\,$s and an inclined magnetic dipole moment $\boldsymbol{\mu}$.
The produced $e^\pm$ pairs fill the closed rotating magnetosphere (shaded in yellow) and form a magnetized relativistic outflow outside of it. The outflow has two zones: the striped wind near the equatorial plane, at polar angles $|\theta-\pi/2|<\chi$, and the helical-$\bB$ wind at $|\theta-\pi/2|>\chi$ \citep{Michel71,Coroniti90}. Far outside the light cylinder, $r\gg\RLC\sim 5\times 10^9\,$cm, the wind magnetic field becomes nearly toroidal, $\bB\approx\bB_\phi$.
The striped wind is shaped by the current sheet (black curve) separating the two regions with opposite magnetic polarities. It gradually dissipates the alternating magnetic fields, converting magnetic energy to heat and then to plasma bulk kinetic energy. The helical-$\bB$  wind is non-dissipative and cold. Its Lorentz factor $\Gw$ grows up to $\sim3\eta^{1/3}$ far outside the magnetosonic radius $R_{\rm ms}\sim \eta^{1/3}\RLC$. We estimate a typical $\eta^{1/3}\sim 5-10$ and  $R_{\rm ms}\simlt 10^{11}$~cm. The thick red curve shows a blast wave launched into the wind by a magnetospheric flare (described in \Sect~\ref{flare}). The blast has Lorentz factor $\Gamma\gg \Gw$. Blast waves in the cold helical-$\bB$ wind are efficient producers of FRBs (\Sect~\ref{maser}).
}
\label{fig:wind}
 \end{figure*}
%%%%%%%%%%% FIGURE %%%%%%%%%%%%%%%%%%

In addition to this cold outflow along the open field-line bundle, the wind is populated at $r\sim\RLC$ with a hot $e^\pm$ component created by the IC photons from the equatorial current sheet (\Eq~\ref{eq:dNrec}). However, the hot component likely makes a minor contribution to the total particle number and energy, compared with the main wind component described by \Eq~(\ref{eq:dN}). Therefore, below we approximate the wind at $r\sim \RLC$ as a mildly relativistic cold outflow emerging from the inner magnetosphere. This approximation should be particularly good at large latitudes, away from the dissipative equatorial current sheet.

The wind flows out through the light cylinder with the subsonic $\gamma<\gamma_s$ and is nearly force-free, as its magnetic energy exceeds the plasma energy by the large factor $\eta>10^2$ (\Eq~\ref{eq:eta}).\footnote{The magnetosonic speed in a fluid with proper mass density $\tilde{\rho}$, magnetization $\sigma=\tilde{B}^2/4\pi \tilde{\rho} c^2$, and enthalpy $w\tilde{\rho} c^2$, is given by \citep{Beloborodov17a}
$$
  \frac{c_s^2}{c^2}=\frac{(\alpha-1)w+\sigma}{1+w+\sigma}, \qquad 4/3<\alpha<5/3.
$$
The corresponding Lorentz factor in the limit of $\sigma\gg 1,w$ is $\gamma_s\approx \sigma^{1/2}[1+(2-\alpha)w]^{-1/2}$. It simplifies to $\gamma_s\approx\sigma^{1/2}$ when $w<1$.}
The standard picture of a magnetic wind from an inclined rotator $\bOm\nparallel\boldsymbol{\mu}$ is summarized in Figure~\ref{fig:wind} \citep{Michel71,Coroniti90}. The cold, helical-$\bB$ zone of the wind will be of main interest for us --- it is the most promising zone for generating FRBs by the shock maser mechanism discussed in \Sect~6 below. 

The wind is accelerated by the pressure of its (dominant) toroidal magnetic field. This pressure is communicated radially by magnetosonic waves, and the wind acceleration saturates after its speed reaches the magnetosonic speed $c_s$ measured in the plasma rest frame (e.g. \cite{Kirk09}). At this magnetosonic point $R_{\rm ms}$ the wind Lorentz factor equals $\eta^{1/3}$ \citep{Michel69,Goldreich70}, and slowly grows to $\sim 3\eta^{1/3}$ at larger radii $r\gg R_{\rm ms}$,
\beq
\label{eq:Gw}
  \Gw\sim 3\eta^{1/3}  \qquad \mbox{(cold~helical-$\bB$~wind)}.
\eeq
The numerical factor in this estimate $\sim 3$ is not exactly constant at $r\gg R_{\rm ms}$ --- it can have a logarithmic dependence on radius  and some dependence on latitude \citep{Lyubarsky2001b}.

The wind acceleration at radii $r<R_{\rm ms}$ and the value of $R_{\rm ms}$ are easy to find when noting that the helical open field lines of a rotating dipole at $r\gg\RLC$ are similar to those of a rotating (split) monopole \citep{Michel73,Bogovalov99}. Their drift velocity $\bv_D/c=(\bE\times\bB)/B^2$ is asymptotically radial, and the drift Lorentz factor $\Gamma_D\approx r/\RLC$ grows linearly between $\RLC$ and $R_{\rm ms}$ until $\Gamma_D$ approaches $\eta^{1/3}$ \citep{Buckley77}. At the magnetosonic radius $R_{\rm ms}$ the fluid inertia becomes important, and the force-free approximation becomes invalid.
Thus, the wind acceleration occurs quickly, approximately linearly with radius, 
until the wind reaches $r\sim \eta^{1/3}\RLC\simlt 10^{11}$~cm, and then $\Gw$ gradually saturates, reaching a few times $\eta^{1/3}$.

The wind magnetization parameter (the ratio of the electromagnetic and kinetic luminosities) is given by
\beq
\label{eq:sigw}
  \sigw=\frac{\Lw}{\Gw\dot{N}m_ec^2}=\frac{\eta}{\Gw}\sim 0.3\,\eta^{2/3}.
\eeq 

The striped wind, which forms around the equatorial plane (Figure~\ref{fig:wind}), is different. Here the magnetic stripes of opposite polarities gradually dissipate, releasing energy up to $\eta m_ec^2$ per particle. The released heat tends to accelerate the equatorial wind \citep{Lyubarsky01,Drenkhahn02}, although recent numerical simulations \citep{Zrake16,Cerutti17b} suggest a more complicated picture of current sheet dissipation. The heated striped wind is a less promising medium for generating FRBs; therefore  in this paper we will focus on the cold helical-$\bB$ zone, where the wind is described by \Eqs~(\ref{eq:Gw}) and (\ref{eq:sigw}).

%###################################################################

\section{Slow ion tails of flares and wind termination shocks}

Magnetar winds have a special feature: they are intermittently polluted by massive ion ejecta from giant flares. This feature becomes important in hyper-active magnetars, which flare frequently.

Observations of SGR~1806-20 suggest that giant flares are capable of ejecting large amounts of plasma, $m_0\simgt 10^{24}$~g, with mildly relativistic speeds $v_0\sim 10^{10}$~cm/s. Its December 2004 flare produced massive ejecta with energy of $\simgt 1$\% of the gamma-ray flare $\E_\gamma$ \citep{Gaensler05,Gelfand05,Granot06}. A significant restructuring of the magnetosphere is required to allow massive ion ejection from the magnetar surface, and hence it must occur on a short timescale $\ti$ comparable to the flare duration.

\subsection{Ion ejecta structure}

At times $t\gg \ti$, the ion outflow may be approximated as ballistically spreading matter from impulsive ejection, with some velocity distribution. It quickly becomes cold due to adiabatic cooling and continues to expand ballistically with a monotonic velocity profile $dv/dr>0$. The outflow structure is determined by the distribution of the ejected mass $m$ over velocity $v$. Let us approximate this distribution as a power-law, which cuts off above some maximum $v_0$ carrying most of the mass,
\beq
\label{eq:m}
   m(v)\equiv \frac{dm}{d\ln v}=m_0\left(\frac{v}{v_0}\right)^\xi, \qquad v\leq v_0,
\eeq 
where $\xi>0$, and $v_0$ is expected to be mildly relativistic, as suggested by the observations of SGR~1806-20. The exact shape of the cutoff at $v>v_0$ is unimportant for the discussion below.

For ejecta with speeds $v$ smaller than the escape velocity from the neutron star, $v_{\rm esc}\sim 0.4c$, gravity may play a role in shaping $v(m)$.\footnote{I thank Yuri Levin for pointing this out to me. Paper~I also argued that gravity limits the mass of the magnetar ejecta. Furthermore, the estimates in Paper~I for the particle number and energy content of the radio nebula around FRB~121102 gave energy per ejected particle comparable to the ion binding energy in the gravitational potential of a neutron star.}
The ejecta tail with $v\ll v_{\rm esc}$ has a small specific energy $E=v^2/2$, so all parts of the tail have approximately the same binding energy $E\approx 0$ when compared with a broad range of $E(m)$ for the matter initially lifted from the star (most of which may have $E<0$ and fall back). Then one can approximate $dm\approx (dm/dE)_{E=0}\,dE\propto dE\propto vdv$ which gives $dm/d\ln v\propto v^{-2}$, i.e. $\xi=2$.

The cold, ballistic ion ejecta extends out to the characteristic radius $R_0(t)=v_0t$ where most of the escaping ion mass is located,
\beq
  R_0 = v_0\trep\sim 3\times 10^{14}\, \frac{v_0}{c}\,\trep_4 {\rm~cm}.
\eeq 
The ejecta has a homologous tail, which occupies a broad range of radii $r=vt\leq R_0(t)$. The mass density distribution in the spreading tail is given by
\beq
\label{eq:rho}
  \rho(r,t)\sim \frac{m(v)}{4\pi r^3}, \qquad v=\frac{r}{t}.
\eeq
Its magnetization parameter $\sigma(v)$ is strongly reduced, as a result of the radial expansion with $dv/dr<0$. It may be roughly estimated as follows. 

Let $\RA$ be the Alfv\'en radius of the outflow during the ion ejection time $\ti$. The outflow is cut off when the flare ends, and we assume that the ejection of all parts of the ion ejecta (with different speeds $v$) occurs on a similar timescale $\ti$.
The magnetic energy ejected through $\RA$ and deposited into mass $m(v)$ is given by
\beq 
   \E_{\rm mag}(v)\sim \BA^2\RA^2 v\ti, 
\eeq
where $\BA(v)$ is the magnetic field deposited in the outflow.
The deposited magnetic flux $\Psi(v)\sim \BA\RA v \ti$ is 
\beq
   \Psi(v)\equiv\frac{d\Psi}{d\ln v} \sim \left[v \ti\, \E_{\rm mag}(v)\right]^{1/2}
         \simlt  \left(mv^3 \ti\right)^{1/2},
\eeq
where we used $\E_{\rm mag}(v)\simlt mv^2$, as expected at the Afv\'en radius.

The deposited flux $\Psi$ is conserved during the homologous expansion of the ion ejecta. At time $\trep\gg \ti$, $m(v)$ is stretched over the radial distance $\sim vt$, and its magnetic field becomes
\beq
  B(v,t)\sim \frac{\Psi}{v^2 t^2}\sim \frac{1}{t^2}\left(\frac{m t_i}{v}\right)^{1/2}.
\eeq 
Using \Eqs~(\ref{eq:rho}) and (\ref{eq:rho}) we find that the magnetization parameter of the ion tail is extremely small,
\beq
\label{eq:sigt}
  \sigma=\frac{B^2}{4\pi\rho c^2}\sim \frac{v^2}{c^2}\,\frac{\ti}{t}.
\eeq

%%%%%%%%%%% FIGURE %%%%%%%%%%%%%%%%%%
\begin{figure*}[t]
% \vspace*{-8cm}
% \hspace*{-0cm}
\begin{center}
\includegraphics[width=0.7\textwidth]{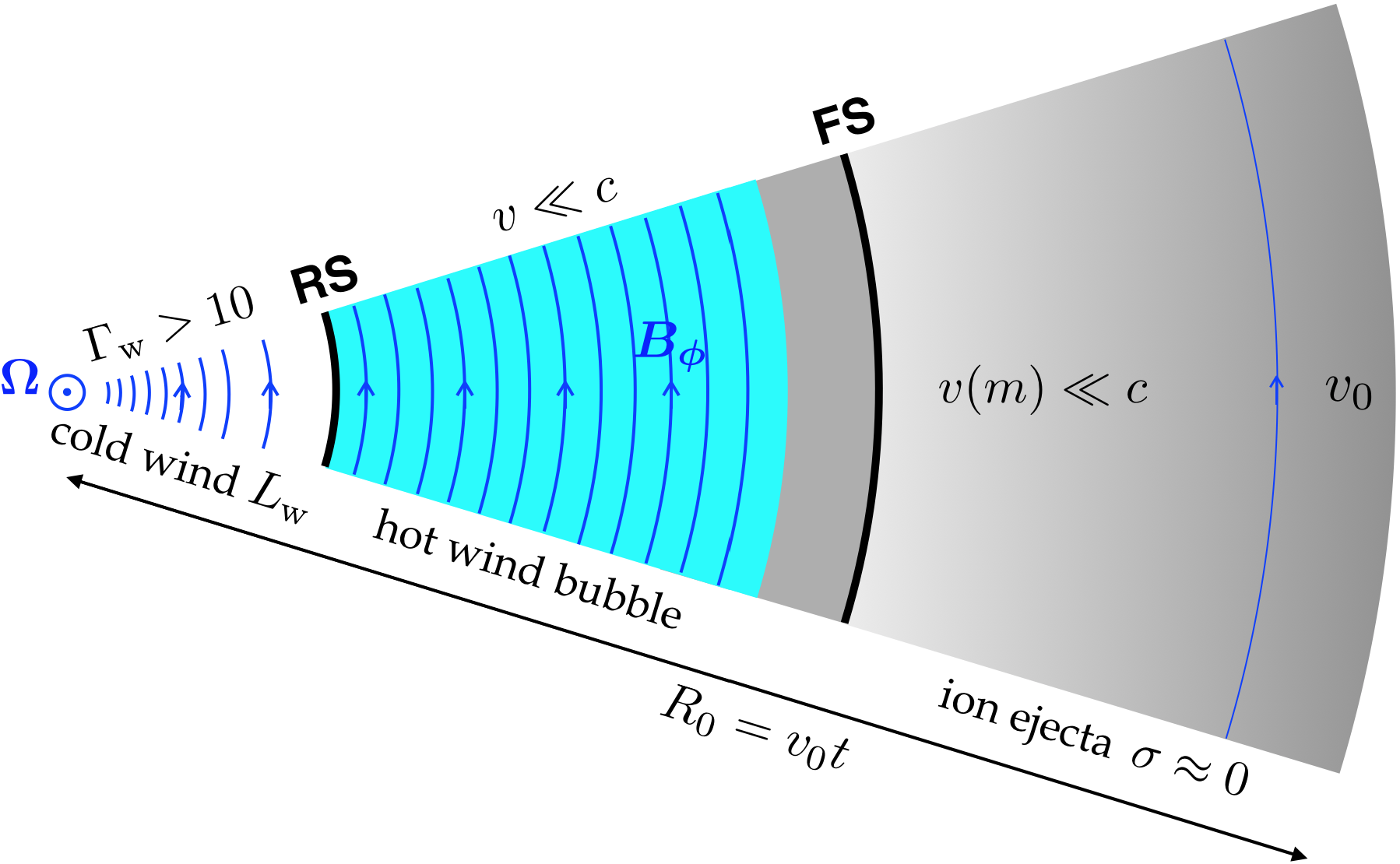} 
\end{center}
% \vspace*{-8.1cm}
\caption{Wind interaction with the tail of ion ejecta. The slow ion matter (grey) was ejected time $t$ ago with a velocity distribution $v(m)\leq v_0$, where $m$ is the Lagrangian (mass) coordinate. The radial spreading $r(m)=v(m)t$ forms the homologous tail of the ion ejecta and reduces its magnetization $\sigma$ to nearly zero (\Eq~\ref{eq:sigt}). The continual wind from the magnetar with power $\Lw$ drives a forward shock (FS) into the ion tail. There is also a reverse shock (RS) at which the wind is decelerated from its Lorentz factor $\Gw\gg 1$ to a sub-relativistic speed.
The RS resembles the wind termination shock in pulsar wind nebulae. Its radius $\RRS$ is estimated in \Eq~(\ref{eq:RRS}). The decelerated wind forms a hot bubble (cyan), with temperature $kT_{\rm b}\sim (\Gw/3)m_ec^2$. The bubble is filled with compressed azimuthal magnetic field of the wind $B_\phi$ (blue arrows). In a complete global picture, the ion ejecta may occupy a moderate solid angle, which allows the wind to flow around it, forming a bow shock. In any case, a hot bubble with pressure comparable to $\Lw/4\pi \RRS^2 c$ forms between the ion ejecta and the fresh, freely expanding wind from the magnetar. 
\label{fig:bubble}
}
 \end{figure*}
%%%%%%%%%%% FIGURE %%%%%%%%%%%%%%%%%%

\subsection{Wind bubble behind the ion ejecta}
\label{bubble}

The ion tail cannot occupy arbitrarily small radii $r\ll R_0(t)$, because there is a persistent spindown wind from the magnetar after the flare. The persistent wind applies pressure $\sim \Lw/4\pi r^2 c$ and sweeps the low-density trailing parts of the ion tail. 

The wind-tail interaction resembles the standard picture of pulsar wind nebulae. It involves a forward shock (FS, propagating in the ion ejecta) and a reverse shock (RS, propagating in the wind), with an approximate pressure balance between the two shocks (Figure~\ref{fig:bubble}). There is also a contact discontinuity between the shocked wind and the shocked ion ejecta. Note that the FS propagates in the ion medium with extremely low magnetization. This shock is mediated by Weibel instability. The RS propagates in the magnetically dominated wind; it is mediated by Larmor rotation. The RS forms because the wind is super-magnetosonic.

The RS is the termination shock of the freely expanding relativistic wind from the magnetar; here the wind decelerates and joins the hot, slowly expanding bubble between the RS and FS, confined by the heavy ion ejecta ahead. The bubble energy equals the energy deposited by the wind during time $\trep$ passed since the ion ejection,
\beq 
\label{eq:Eb}
   \Eb\approx \int_0^{\trep} \Lw\, dt^\prime 
   =10^{42}\bar{L}_{\rm w,38}\, t_4 {\rm~erg},
\eeq
where $\bar{L}_{\rm w}$ is the average power of the spindown wind during time $t$.

Let $\RRS$ and $\RFS$ be the radii of the reverse and forward shock, respectively. Approximating the bubble volume as $V_b\sim (4\pi/3)\RFS^3$ and the bubble pressure $P$ as the pressure at the RS,
\beq
  P\sim \frac{\Lw}{4\pi\RRS^2 c}, 
\eeq
one finds a relation between $\RRS$ and $\RFS$ from $3PV_b\sim \Eb$,
\beq
\label{eq:R}
   \frac{\RFS^3}{c\RRS^2}\sim t.
\eeq
The ratio of the dissipation rates in the FS and RS is $\RFS^2\vFS/\RRS^2 c\sim \RFS^3/\RRS^2ct\sim 1$. Thus, the deposited wind energy $\Eb$ is roughly equally partitioned between the wind (heated and compressed in the RS) and the ion plasma (heated and compressed in the FS). 

The FS propagation in the homologous ion ejecta with the velocity distribution~(\ref{eq:m}) approximately satisfies
\beq
   \vFS-v\sim v,
\eeq
and the energy dissipated in the FS, $\sim \Eb/2$, is comparable to the upstream kinetic energy, so
\beq
   m(v) v^2\sim \Eb.
\eeq
This gives an estimate for the mass $m$, and the corresponding speed $v(m)$, of the ions at the FS location,
\beq
   \frac{m}{m_0}\sim\left(\frac{\Eb}{\Et}\right)^{\xi/(2+\xi)}<1,
\eeq
where $\Et\sim m_0 v_0^2$ is the total energy of the ion ejecta. In the December 2004 flare of SGR~1806-29, the ejecta was estimated to carry a large $\Et\simgt 10^{44}$~erg \citep{Gelfand05,Granot06}; therefore we will assume $\Et\gg \Eb$. 

The forward shock radius $\RFS\sim \vFS t\sim v t$ is then given by  
\beq
\label{eq:RFS}
  \RFS\sim R_0 \left(\frac{\Eb}{\Et}\right)^{1/(2+\xi)}.
\eeq
Combining with \Eq~(\ref{eq:R}) we also find
\beq
  \frac{\RRS}{\RFS}\sim \left(\frac{v_0}{c}\right)^{1/2} \left(\frac{\Eb}{\Et}\right)^{1/(4+2\xi)},
\eeq
\beq
   \RRS\sim R_0\left(\frac{v_0}{c}\right)^{3/2} \left(\frac{\Eb}{\Et}\right)^{3/(4+2\xi)}.
\eeq
One can see that the radius of the wind termination shock (RS) $\RRS$ rather weakly depends on the ejecta velocity distribution, which is described by the index $\xi$. 
In particular, for $\xi=2$ we find
\beq
\label{eq:RRS}
     \RRS\sim 10^{14} \left(\frac{v_0}{0.3c}\right)^{3/2}  \left(\frac{t}{1\,\rm d}\right)^{11/8} \bar{L}_{\rm w,38}^{3/8}\,\E_{\rm t,44}^{-3/8}{\rm ~cm}.
\eeq
Practically the same $\RRS$ is found when $\xi=1$, with the scaling exponents $11/8$ and $1/2$ changed to $3/2$ and $3/8$. 
As the spindown wind with a Lorentz factor $\Gw$ passes through the RS, it decelerates to $\sim v$ and its plasma kinetic energy transforms into heat. Thus, the slowly expanding bubble of the terminated wind outside radius $\RRS$ has a high temperature, with a thermal Lorentz factor $\gamma_b\sim \Gw$. 

Note that the above description of the wind bubble assumed, for simplicity, quasi-isotropic ion ejecta. In a more realistic picture, the ejecta occupy a limited solid angle (as observed in SGR~1806-20), and form a blob rather than a spherical shell. Then the magnetar wind may not be confined by the blob. Instead, the wind may flow around it and form a bow shock. The hot bubble behind the ion ejecta (now confined by the bow shock) will retain less energy than $\Eb$ estimated in \Eq~(\ref{eq:Eb}).

%#################################################################

\section{Blast waves from magnetic flares}
\label{flare}

Magnetospheric flares drive explosions into the magnetar wind, resembling shocks in the solar wind launched by solar flares. In contrast to the solar activity, the winds and explosions from magnetars are relativistic.

\subsection{Ejection of ultra-relativistic plasmoids}

The basic mechanism of magnetar flares resembles that of solar flares: strongly twisted magnetic loops inflate, reconnect, and eject plasmoids. The biggest plasmoids can carry a significant fraction of the magnetic loop energy. The dynamics of magnetar flares is demonstrated by force-free electrodynamic simulations \citep{Parfrey13,Carrasco19}. 

The plasmoid energy can be quite enormous, especially if the magnetar has spots of concentrated magnetic field, resembling sunspots. A loop with $B\sim 10^{16}$~G occupying a fraction $f$ on the stellar surface, and extending to a scale-height comparable to the stellar radius $R_\star$, has energy $\sim f R_\star^3 B^2/8\pi\approx  4\times 10^{46}\,f_{-2} B_{16}^2$~erg. It is capable of ejecting a plasmoid of energy up to $\E_{\rm pl}\sim 10^{46}$~erg.

A gamma-ray flare with energy $\E_\gamma\sim 10^{46}$~erg was observed from SGR~1806-20 in December 2004  \citep{Palmer05}. The flare also produced mildly-relativistic ejecta with energy $E>10^{44}$~erg \citep{Gelfand05,Granot06}. No ejection of ultra-relativistic plasmoids were reported, and their detection may be difficult, because their emission was likely beamed away from our line of sight. Theoretically, the ratio $\E_{\rm pl}/\E_\gamma$ can strongly vary, and could in principle be much smaller or larger than unity.

The plasmoid ejected by an over-twisted magnetic loop will be utra-relativistic, because its baryon loading is low and its $e^\pm$ content is limited by $e^\pm$ annihilation (Paper~I).  
The plasmoid size is comparable to the loop size in the inner magnetosphere, $\Delta\sim 10^7$~cm. It begins to expand as it accelerates away from the star, and may occupy a moderate solid angle $\Omega_{\rm pl}\ll 4\pi$. Hereafter $\E$ will denote the apparent {\it isotropic equivalent} of the plasmoid energy,
\beq
   \E\sim \left(\frac{4\pi}{\Omega_{\rm pl}}\right)\,\E_{\rm pl}.
\eeq 

At radii $r\gg 10^7$~cm, the plasmoid is basically a shell of concentrated magnetic energy flying radially away from the neutron star. Dynamics of such shells in vacuum and static external media was calculated by \cite{Lyutikov10}, \cite{Levinson10}, and \cite{Granot11}. The shell thickness $\Delta$ remains approximately constant, while its transverse dimension grows proportionally to radius, and the plasmoid soon looks like a thin pancake. Its Lorentz factor grows as
\beq
   \Gf(r)\approx \left(\frac{\eta_{\rm f}\, r}{\Delta}\right)^{1/3}
     = 10^5\,r_{13}^{1/3}\,\eta_{\rm f,9}^{1/3}\,\Delta_7^{-1}.
\eeq
Here $\eta_{\rm f}=\E/\N m_ec^2$ and $\N$ is the isotropic equivalent of the $e^\pm$ number carried by the $\Delta$-shell. 

The estimate of $\eta_{\rm f}$ was made in Paper~I, and is repeated below for completeness. Since the $\Delta$-shell is ejected at the beginning of the flare, it does not have time to become loaded with baryons lifted from the magnetar surface.  It may initially be loaded with a hot $e^\pm$ plasma, which annihilates later as the $\Delta$-shell expands and cools. The freezes-out number of surviving pairs corresponds to Thomson optical depth $\tau_{\rm T}\sim 1$, because the annihilation cross section is comparable to Thomson cross section $\sT$. The $e^\pm$ freeze-out occurs where the shell temperature drops to $kT_\pm\sim 20$~keV, at radius $R_\pm\sim 10^8$~cm. The condition $\tau_{\rm T}\sim 1$ implies $e^\pm$ density $n_\pm\sim \Gamma_\pm^2 / R_\pm\sT$ where $\Gamma_\pm\sim 10$ is the shell Lorentz factor after adiabatic cooling from the initial temperature of a few hundred keV to the freeze-out temperature $kT_\pm\sim 20$~keV. The resulting $\N$ may be roughly estimated as $\N\sim 4\pi \Gamma_\pm^2  R_\pm \Delta/\sT\sim 10^{42}$. 

The large value of $\eta_{\rm f}\sim 10^8\,\E_{\rm 44}\N_{42}^{-1}$ gives so high $\Gf$ that its exact value becomes irrelevant. The results of this paper would remain the same for $\eta_{\rm f}\sim 10^5$. The Lorentz factor $\Gf$ may also be taken as infinite at radii of interest, and then the $\Delta$-shell can be viewed as an electromagnetic pulse propagating with the speed of light \citep{Lyubarsky14}. The hyper-relativistic plasmoid ejected from the over-twisted magnetic loop is followed by a slower outflow from the giant flare. The outflow likely has both an ultra-relativistic component (which emits the observed gamma-ray peak with $\sim 0.5$~s duration \citep{Thompson01}), and a mildly relativistic component (which emits the radio afterglow of the December 2004 flare of SGR~1806-20).

The leading plasmoid of thickness $\Delta$ acts as a piston driving a blast wave into the ambient medium, i.e. into the pre-explosion wind from the magnetar (Figure~\ref{fig:wind}). The  piston motion relative to the wind has Lorentz factor $\Gf/2\Gw$. This motion is super-magnetosonic (the magnetosonic waves in the wind have Lorentz factor $\gamma_s\approx \sigw^{1/2}$), and so it launches a strong shock into the wind. As discussed below, the resulting blast wave has Lorentz factor $\Gamma\simgt 10^3$ until it decelerates. Below we focus on the ultra-relativistic stage of the explosion, $\Gamma\gg 10$. We will assume that any ejecta moving behind the $\Delta$-shell have a lower Lorentz factor and thus do not contribute to the blast wave at its early deceleration stage.

\subsection{Blast wave dynamics in the wind}
\label{bw}

It takes time for the plasmoid ($\Delta$-shell) to transfer all its energy to the blast wave ahead of it. The radius at which the energy transfer is accomplished was denoted by $R_{\rm tr}$ in Paper~I; hereafter we change its notation to $\Rdec$. Quantities evaluated at $\Rdec$ will have subscript $\diamond$. Before reaching $\Rdec$, the blast wave energy $\Ebw$ grows linearly with radius while its Lorentz factor $\Gamma$ remains approximately constant.

%%%%%%%%%%% FIGURE %%%%%%%%%%%%%%%%%%
\begin{figure}[t]
% \begin{center}
\includegraphics[width=0.46\textwidth]{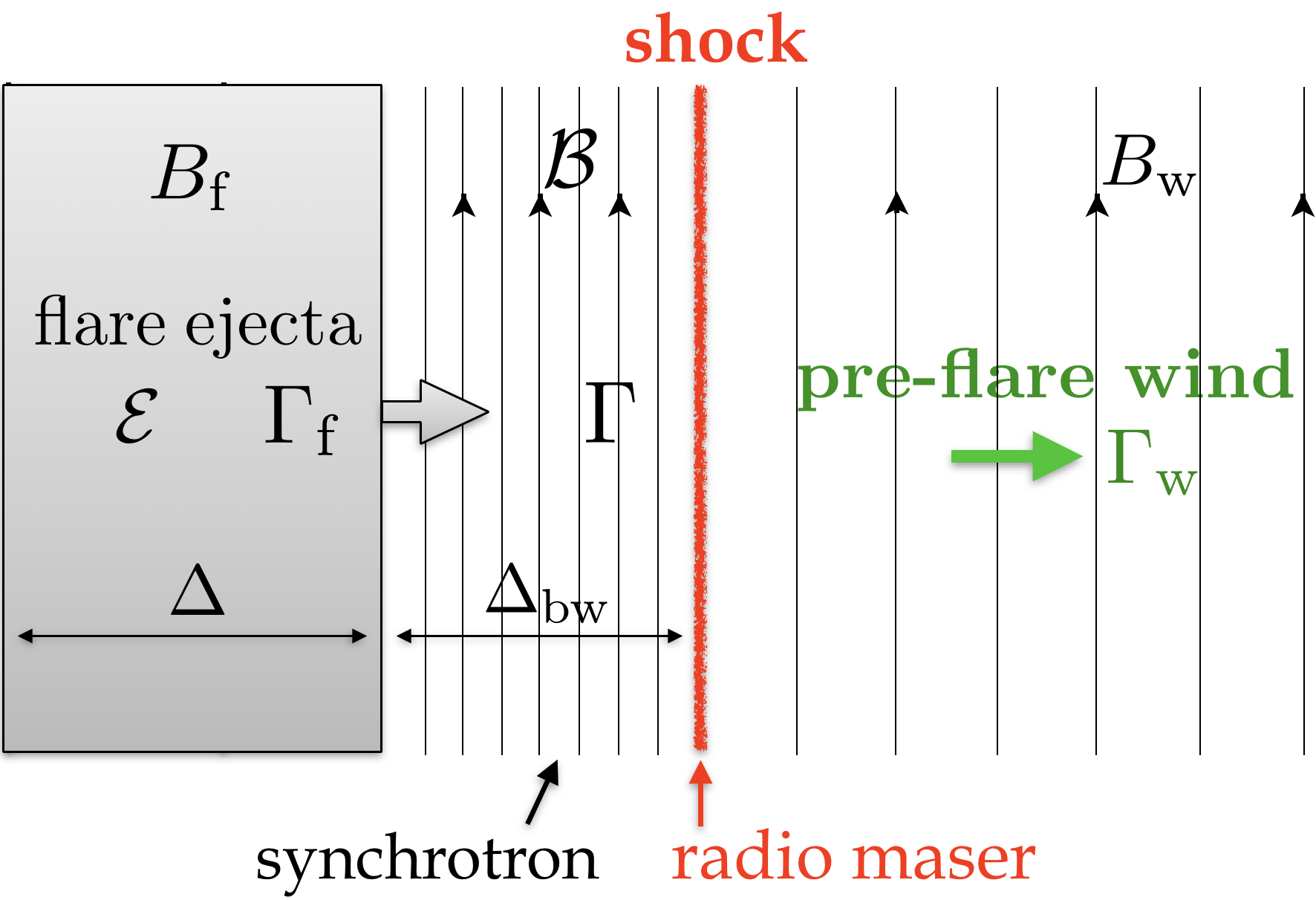} 
% \end{center}
\caption{Ultra-relativistic blast wave (Lorentz factor $\Gamma\simgt 10^3$) is driven into the magnetar wind ($\Gw\simgt 10$) by the plasmoid ejected during a giant flare of the magnetar. The plasmoid has a typical energy $\E\sim 10^{44}$~erg, thickness $\Delta\sim 10^7$~cm and a Lorentz factor $\Gf\gg\Gamma$; it acts like a piston driving the forward shock (red) into the magnetized wind. The shock is mediated by Larmor rotation, which forms an unstable soliton-type structure, continually generating strong, semi-coherent electromagnetic waves (radio maser) described in \Sect~6. The thermalized particles behind the shock have Lorentz factors $\gamma_{\rm th}\sim\Grel=\Gamma/2\Gw$. They emit synchrotron radiation in the compressed magnetic field $\B$. This emission is too weak to be observed unless the blast wave strikes the tail (wind bubble) accumulated behind massive, slow ion ejecta of a previous flare; then a bright optical flash will be produced, as described in \Sect~5.
}
\label{fig:shock}
 \end{figure}
%%%%%%%%%%% FIGURE %%%%%%%%%%%%%%%%%%

The explosion structure at $r<\Rdec$ is shown in Figure~\ref{fig:shock}.
Following Paper~I, we define the ``flare power'' $\Lf$ by 
\beq
  \Lf=\frac{\E}{\tau}=10^{47}\,\frac{\E_{44}}{\tau_{-3}}\; \frac{\rm erg}{\rm s}, 
   \qquad \tau\equiv\frac{\Delta}{c}.
\eeq 
The swept-up wind material ahead of the plasmoid forms a ``blast'' 
--- a shell of thickness $\Delta_{\rm bw}\sim r/\Gamma^2$ with a nearly uniform Lorentz factor $\Gamma$, which is determined by the approximate pressure balance with the plasmoid. Pressure is everywhere magnetically dominated, and so pressure balance implies the approximate equality between the magnetic field in the plasmoid,
\beq
\label{eq:Bf}
    \Bf\approx \left(\frac{\Lf}{c\,r^2}\right)^{1/2},
\eeq
and the compressed wind field in the blast wave,
\beq
\label{eq:B}
   B \approx \frac{\Gamma^2}{\Gw^2}\,\Bw
   \approx \frac{\Gamma^2}{\Gw^2}\left(\frac{\Lw}{c\,r^2}\right)^{1/2}.
\eeq
Both $B$ and $\Bf$ are measured in the static lab frame. The balance $B\approx\Bf$ gives
\beq
\label{eq:G1}
   \Gamma\approx\Gw\left(\frac{\Lf}{\Lw}\right)^{1/4}  \qquad (r<\Rdec).
\eeq 

The blast thickness in the lab frame is 
\beq
\label{eq:dbw}
  \dbw\sim \frac{r}{\Gamma^2},
\eeq 
and its energy (isotropic equivalent) is 
\beq
\label{eq:Ebw}
  \Ebw\sim 4\pi r^2 \dbw \frac{B^2}{8\pi}\sim \frac{r^3 B^2}{2\Gamma^2}.
\eeq
At $r<\Rdec$ one can use $B\approx \Bf$ to rewrite \Eq~(\ref{eq:Ebw}) as 
\beq
\label{eq:Ebw1}
   \Ebw\sim \Lf\,\frac{r}{2\Gamma^2 c}  \qquad (r<\Rdec).
\eeq

The transfer of the plasmoid energy $\E$ to the blast wave is accomplished where $\Ebw$ approaches $\E$. This condition gives
\begin{eqnarray}
\nonumber
   \Rdec &\sim& 2\Gamma^2 c\tau \approx 2\Gw^2\,c \left(\frac{\tau\E}{\Lw}\right)^{1/2} \\
   &\approx & 2\times 10^{14}\,\left(\frac{\Gw}{20}\right)^2
   \left(\frac{\tau_{-3}\,\E_{44}}{L_{\rm w,39}}\right)^{1/2} {\rm cm},
\label{eq:Rdec}
\end{eqnarray}
where we substituted \Eq~(\ref{eq:G1}) for $\Gamma$. 

%%%%%%%%%%% FIGURE %%%%%%%%%%%%%%%%%%
\begin{figure}[t]
% \begin{center}
\includegraphics[width=0.46\textwidth]{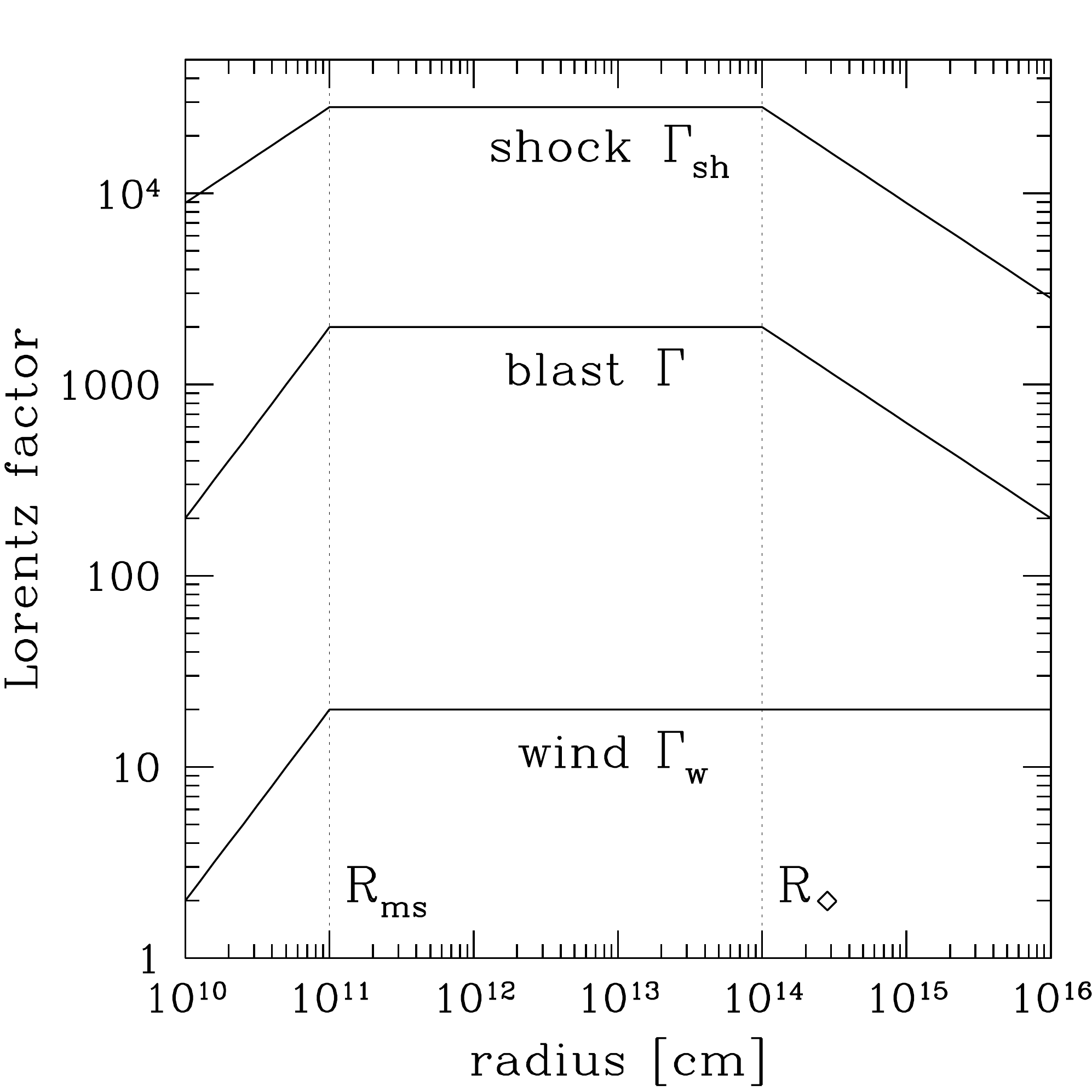} 
% \end{center}
\caption{Evolution of the expanding blast wave in the magnetar wind with $\Lw\approx 10^{39}$~erg/s and $\eta\approx 10^3$. The explosion energy in this example is $\E\approx 10^{44}$~erg. Three Lorentz factors are shown: $\Gw$ (the pre-explosion wind), $\Gsh$ (the shock), and $\Gamma$ (the blast --- the hot plasma behind the shock). The shock moves faster than the blast: $\Gsh\approx 2\sigw^{1/2}\Gamma$ according to the jump conditions for ultra-relativistic shocks in a strongly magnetized plasma, $\sigw\gg 1$. The wind magnetization is determined by the relation $\Gw\sigw=\eta$. The wind Lorentz factor outside the magnetosonic radius $R_{\rm ms}$ was approximated here as constant $\Gw\approx 2\eta^{1/3}$; in a more detailed model $\Gw$ would slowly, as $(\ln r)^{1/3}$, grow at $r\gg R_{\rm ms}$ up to $\Gw\sim 3\eta^{1/3}$ at $\Rdec$. The blast wave begins to decelerate outside the radius $\Rdec$ given in \Eq~(\ref{eq:Rdec}).
}
\label{fig:bw}
 \end{figure}
%%%%%%%%%%% FIGURE %%%%%%%%%%%%%%%%%%

The evolution of $\Gamma$ at $r>\Rdec$ is found by setting $\Ebw=\E$ in \Eq~(\ref{eq:Ebw}) and using \Eq~(\ref{eq:B}) for $B$,
\begin{eqnarray}
\nonumber
    \Gamma  &\approx& \Gw^2\left(\frac{2c\,\E}{r\Lw}\right)^{1/2}  \quad\quad (r>\Rdec) \\
     &\approx& 3\times 10^3 \, \left(\frac{\Gw}{20}\right)^2 \frac{\E_{44}^{1/2}}{r_{14}^{1/2}L_{\rm w,39}^{1/2}}. 
\label{eq:G2}
\end{eqnarray}
For explosions in a steady wind with $\Gw\approx const$, one finds $\Gamma\propto r^{-1/2}$ i.e. $\Gamma$ decreases at $r>\Rdec$. Thus, the transfer radius $\Rdec$ is also where the blast wave begins to decelerate. The blast wave evolution in the magnetar wind is shown in Figure~\ref{fig:bw}.

\subsection{Blast wave impact on the slow tail of a previous flare}
\label{impact}

Hyper-active magnetars may have periods of frequent flaring, with recurrence times $t$ as short as days or perhaps even minutes, in extreme cases. Such behavior is suggested by the observed periods of extremely frequent bursting of FRB~121102. The blast wave from each flare will propagate in the free wind emitted by the spinning magnetar ahead of the flare. The blast wave may also impact the slow tail of a previous flare, if there is such a tail (only sufficiently strong flares may be capable of having massive ion tails). Where this impact happens depends on the time $t$ separating the two consecutive flares.

The slow tail described in \Sect~\ref{bubble} will now be considered as a target or obstacle for the ultra-relativistic blast wave. The blast wave reaches the obstacle where the free wind ends and the hot bubble begins, i.e. at radius  $\RRS$ (the wind termination shock marked as RS in Figure~\ref{fig:bubble}). This radius is given in \Eq~(\ref{eq:RRS}).

At radius $\RRS$ the upstream medium suddenly changes from the cold wind with Lorentz factor $\Gw$ and power $\Lw$ to the hot, slow bubble with pressure $P\approx \Lw/4\pi \RRS^2 c$. As the blast wave crosses $\RRS$ its Lorentz factor suddenly drops by the factor of $\Gw$.

Two cases are possible:
\\
(1) The impact on the bubble occurs before all of the plasmoid energy is transferred to the blast wave, i.e. $\RRS<\Rdec$. Then, at the time of the impact, the blast has magnetic field $B\approx B_f$ (\Eq~\ref{eq:Bf}), and its Lorentz factor $\Gamma$ jumps at $\RRS$ from the value given by \Eq~(\ref{eq:G1}) to 
\beq
\label{eq:G0a}
  \Gamma_0\approx \left(\frac{\Lf}{\Lw}\right)^{1/4}\approx 10^2\,\frac{L_{\rm f,47}^{1/4}
  }{L_{\rm w,39}^{1/4}}.
\eeq
\\
(2) The other possible case is $\RRS>\Rdec$. Then at the impact time the blast already carries the entire plasmoid energy $\E$, and its Lorentz factor jumps at $\RRS$ from the value given by \Eq~(\ref{eq:G2}) to 
\beq
\label{eq:G0b}
  \Gamma_0\approx \Gw\,\left(\frac{c\,\E}{\RRS\Lw}\right)^{1/2}
    \approx 10^2 \, \left(\frac{\Gw}{20}\right) \frac{\E_{44}^{1/2}}{\RRS_{14}^{1/2}L_{\rm w,39}^{1/2}}.
\eeq

\begin{figure}[t]
% \begin{center}
\includegraphics[width=0.46\textwidth]{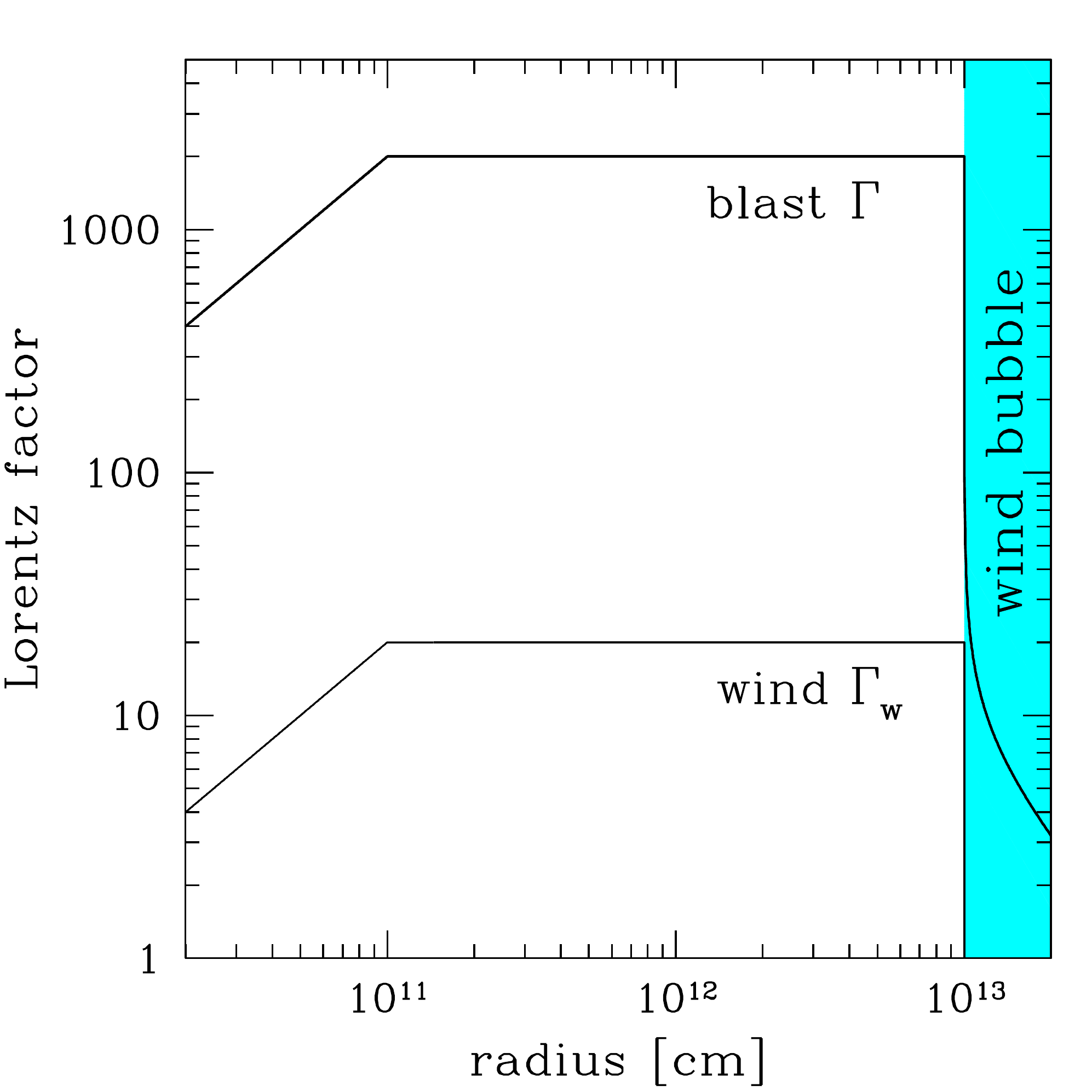} 
% \end{center}
\caption{Blast wave striking the slow tail of a previous flare. The wind and the explosion parameters are the same as in Fig.~\ref{fig:bw}, however now the free wind ahead of the blast wave is terminated at radius $\RRS=10^{13}$~cm and there is a hot wind bubble outside $\RRS$ with $\Gw\sim 1$ (see Fig.~\ref{fig:bubble}). The blast wave steeply decelerates when it enters the bubble and emits a synchrotron optical flash described  in \Sect~\ref{bwb}.
}
\label{fig:bw_bubble}
 \end{figure}

The subsequent evolution of $\Gamma(r)$ in the bubble is determined by the swept-up volume,
 \beq
 \label{eq:V}
   V(r)=\frac{4\pi}{3}  \left(r^3-\RRS^3\right).
 \eeq
The swept-up enthalpy $\sim 4PV$ is boosted by the blast wave by the factor of $\Gamma^2$, and so the energy deposited in the bubble is $\sim 4PV\Gamma^2$. During an initial impact stage with $\Gamma\approx\Gamma_0$ the deposited energy grows linearly with $V$ until it approaches the total explosion energy $\E$. This happens when $V(r)$ reaches 
\beq
\label{eq:V0}
    V_0\sim  \frac{\E}{4P\Gamma_0^2}.
\eeq
Here $P\sim \Lw/4\pi r^2 c$ is the bubble pressure (\Sect~\ref{bubble}).
For the typical parameters of explosions discussed in this paper one finds that $V_0$ is small in the sense that $V_0=4\pi\RRS^2 l_0$ with $l_0\ll\RRS$. Indeed, one finds from \Eq~(\ref{eq:V0}),
\beq
   l_0\sim \frac{c\, \E}{4 \Lw \Gamma_0^2}
   \sim 10^{11}\,\frac{\E_{44}}{L_{\rm w,39}} \left(\frac{\Gamma_0}{100}\right)^{-2} {\rm cm} \ll \RRS.
\eeq
Hence, the explosion energy $\E$ becomes deposited in the slowed blast wave in the bubble almost immediately following the impact. After that, the blast wave energy remains equal to $\E$ (with negligible radiative losses) and $\Gamma$ is determined by energy conservation $4PV\,\Gamma^2 \sim \E$. The blast wave evolution in the bubble may be summarized as follows
\begin{eqnarray}
  \Gamma(r)\approx \Gamma_0\left\{\begin{array}{ll}
                               1          & \qquad V(r)<V_0 \\
                    (V/V_0)^{-1/2} & \qquad V(r)>V_0.
                                                          \end{array}\right.
\label{eq:Gb}
\end{eqnarray}
The resulting evolution is shown in Figure~\ref{fig:bw_bubble} for the case when the two flares were separated by time $t\sim 10^4$~s.

%%%%%%%%%%% FIGURE %%%%%%%%%%%%%%%%%%
%%%%%%%%%%% FIGURE %%%%%%%%%%%%%%%%%%

When the blast wave crosses the bubble and enters the ion ejecta its Lorentz factor will have dropped to 
\beq
\label{eq:Gb1}
  \Gamma\sim (\E/\E_{\rm b})^{1/2}=10\; \E_{\rm 44}^{1/2}\, \E_{\rm b,42}^{-1/2}.
\eeq 
Now the upstream medium is denser and colder; it has the electron-ion (instead of $e^\pm$) composition and a much lower magnetization $\sigma$.\footnote{Some magnetic fields had been generated in the ion ejecta by the FS of the wind-tail interaction shown in Figure~\ref{fig:bubble}. However shock-generated fields tend to quickly decay behind the FS.}
If $\sigma<10^{-3}$, the blast wave becomes mediated by Weibel instability rather than by Larmor rotation.

%####################################################################
 
% \newpage
 
\section{Optical flash}
\label{optical}

Next, we consider emission produced by the blast wave. This section examines incoherent synchrotron emission. Radio maser emission will be discussed in \Sect~\ref{maser}.

The blast wave in the magnetically dominated wind dissipates only a fraction $\sigw^{-1}\ll 1$ of its energy $\Ebw$. Therefore, its dynamics may be approximated as adiabatic, regardless of the plasma radiative losses. However, the radiated energy fraction becomes interesting when one would like to know the observed luminosity. In the radio band, the blast wave emits $\sim 10^{-5}-10^{-6}$ of its energy (see \Sect~\ref{maser} below). A larger fraction can be emitted in synchrotron photons, in particular in the optical band, as shown in \Sect~\ref{bwb} below.

\subsection{Synchrotron emission}

A blast with Lorentz factor $\Gamma\gg 1$ has a characteristic thickness $\delta r\sim r/\Gamma^2$. Its energy $\Ebw$ is related to the magnetic field $\B$ measured in the blast rest frame by $\Ebw\sim r^2\delta r\,\Gamma^2\B^2/2$. This gives 
\beq
\label{eq:Bc}
   \B \sim\left(\frac{2\Ebw}{r^3}\right)^{1/2}
   \sim 400\,\frac{\E_{\rm bw,44}^{1/2}}{r_{13}^{3/2}} {\rm~G}.
\eeq
Note that $\B$ is independent of $\Gamma$ and $\Gw$.

The plasma in magnetized blast waves forms a thermal (Maxwellian) distribution behind the shock \citep{Langdon88,Hoshino91}. Let $\gth$ be the thermal Lorentz factor. The synchrotron cooling timescale in the blast frame, $\tc$, should be compared with the expansion timescale, $\texp$,
\beq
  \tc=\frac{6\pi m_ec}{\sT \B^2\gth},   \qquad \texp=\frac{r}{\Gamma c}.
\eeq
Their ratio is 
\beq
\label{eq:eff}
   \frac{\tc}{\texp} \sim \frac{3\pi m_ec^2 r^2  \Gamma}{\sT \Ebw\gth}. 
\eeq
When $\tc>\texp$ (the ``slow-cooling'' regime) only a fraction $\sim \tc/\texp$ of the dissipated energy is radiated. 
The synchrotron luminosity peaks at the observed frequency (measured in the static lab frame),
\beq
\label{eq:nus}
  \nu_s \sim \frac{0.3e\B }{2\pi m_ec}\,\gth^2\Gamma. 
\eeq

\subsection{Blast wave in the cold wind}

When the cold $e^\pm$ wind with Lorentz factor $\Gw$ is swept-up by the blast wave with Lorentz factor $\Gamma\gg\Gw$, it is heated to the thermal Lorentz factor 
\beq
  \gth=\Grel\approx \frac{\Gamma}{\Gw(1+\beta_{\rm w})}.
\eeq
At $r<\Rdec$
the pressure balance requires $\Gamma\B\approx \Bf=(\Lf/cr^2)^{1/2}$ and $\Grel\approx\Gamma/2\Gw\approx 50(L_{\rm f,47}/L_{\rm w,39})^{1/4}$ (\Sect~\ref{bw}). This gives
\beq
\label{eq:nus1}
  \nu_s\sim \frac{4\times 10^{14}\,{\rm Hz}}{r_{13}}\,\frac{L_{\rm f,47}}{L_{\rm w,39}^{1/2}} 
  \qquad (r<\Rdec).
\eeq
Note that $\nu_s$ is independent of $\Gw$ and close to the optical band for radii around $ 10^{13}$~cm. Using \Eq~(\ref{eq:Ebw1}) for $\Ebw$, we also find 
\beq
\label{eq:eff1}
   \frac{\tc}{\texp} \sim \frac{\Gw^3 r_{13}}{(L_{\rm f,47}L_{\rm w,39})^{1/2}}\gg 1
    \qquad (r<\Rdec).
\eeq
We conclude that the optical flash from the blast wave in an ultra-relativistic cold wind $\Gw\gg 1$ will be weak, because $\tc\gg\texp$. The shocked plasma will lose almost all its energy to adiabatic cooling instead of emitting it. The same is true at larger $r>\Rdec$.

\subsection{Impact on the tail of a previous flare}
\label{bwb}

A bright optical flash can be emitted if the blast wave impacts the wind bubble in the tail of a previous flare, as described in \Sect~\ref{impact}. The impact occurs at radius $\RRS$, the blast wave immediately decelerates to $\Gamma_0\sim 10^2$, and soon decelerates to $\Gamma\simlt 10$. Its evolution in the bubble is given by \Eq~(\ref{eq:Gb}) and shown in Figure~\ref{fig:bw_bubble}.

The bubble is hot, with a thermal Lorentz factor $\gb\sim\Gw$, before the blast wave arrives. The blast wave heats the bubble by the additional factor of $\Gamma$ to
\beq
   \gth\approx \Gamma\gb\sim\Gamma\Gw,
\eeq
and also gives the hot plasma the bulk Lorentz factor $\Gamma$.
The peak frequency of its synchrotron emission (\Eq~\ref{eq:nus}) then becomes,
\beq
\label{eq:nus2}
  \nu_s \sim \frac{0.3\,e\,\E^{1/2}\Gamma^3\gb^2 }{2\pi m_ec\, r^{3/2}}
  \sim \frac{10^{15}\,{\rm Hz}}{\RRS_{14}^{3/2}} \left(\frac{\Gamma}{30}\right)^3
   \left(\frac{\gb}{20}\right)^2\,\E_{46}^{1/2}.
\eeq
Here we normalized $\Ebw\approx \E$ to a high value of $10^{46}$~erg, because strongest explosions give the brightest emission. The blast wave decelerates from $\Gamma_0\sim 10^2$ to $\Gamma$ of a few tens in a narrow range of radii $r-\RRS<\RRS$  (\Sect~\ref{impact}); therefore we substituted $r\approx\RRS$ in \Eq~(\ref{eq:nus2}). During this deceleration, $\nu_s\propto\Gamma^3\propto (r-\RRS)^{-9/2}$ sweeps through a broad frequency range, passing through the optical band. 

The characteristic arrival time of radiation emitted by a blast wave with Lorentz factor $\Gamma$ is $\tobs\sim r/\Gamma^2 c$. The same combination of $r$ and $\Gamma$ appears in \Eq~(\ref{eq:nus2}) for $\nu_s$, and so the observed evolution of the synchrotron peak frequency is given by
\beq
\label{eq:nus3}
   \nu_s(\tobs) \sim \frac{10^{14}\,{\rm Hz}}{\tobs^{3/2}}\left(\frac{\gb}{20}\right)^2\,\E_{46}^{1/2},
\eeq
where $\tobs$ is in seconds.

The radiative cooling efficiency for the blast wave in the bubble is found using \Eq~(\ref{eq:eff}) with $\gth\approx \Gamma\gb$,
\beq
\label{eq:eff2}
  \frac{\tc}{\texp} \sim\frac{6\pi m_ec^2 \RRS^2}{\sT \E \gb}   
  \sim 1\, \RRS_{14}^2 \left(\frac{\gb}{20}\right)^{-1} \E_{\rm 46}^{-1}.
\eeq
This shows that a significant fraction of the dissipated energy $\E_{\rm diss}\sim\E/\sigw$ may be radiated, if the impact occurs at radius $\RRS\sim 10^{14}$~cm. The radiated fraction  is independent of $\Gamma$ and remains approximately constant as the blast wave steeply decelerates at $r\approx\RRS$.

The steep sweeping of $\nu_s(r)$ at $r-\RRS\ll \RRS$ with constant radiative efficiency implies that the decelerating blast wave emits comparable energies in different frequency bands, $d\E_s/d\ln\nu \approx const$. Therefore, the peak luminosity at frequencies $\sim\nu$ may be estimated as 
\beq
\label{eq:L}
   L\sim \frac{1}{\tobs}\,\frac{\texp}{\tc}\,\frac{\E}{\sigw}
   \sim 10^{45}\, \frac{ \nu_{15}^{2/3}\, \E_{46}^{5/3}}{\RRS_{14}^2\,\eta_3^{5/9}} \;\frac{\rm erg}{s},
\eeq
where we substituted $\gb\sim\Gw\sim 3\eta^{1/3}$ and $\sigw=\eta/\Gw$.
The estimate~(\ref{eq:L}) is valid if $\RRS>10^{14}\,\E_{46}^{1/2}\eta_3^{1/6}\,$cm. In the opposite case, $\tc<\texp$, $L$ is a large fraction of the dissipated power,
\beq
   L\sim \frac{\E}{\eta^{2/3}\tobs}\sim \frac{10^{44}\,{\rm erg}}{\tobs}\,\frac{\E_{46}}{\eta_3^{2/3}}   
   \qquad   (\RRS_{14}<\E_{46}^{1/2}\,\eta_3^{1/6}).
\eeq 

We conclude that the optical flash energy can be comparable to the upper limit,
\beq
\label{eq:L1}
  \E_O\simlt \frac{\E}{\sigw},
\eeq 
where $\sigw\sim\eta/\Gw$ is the magnetization parameter of the bubble, which is comparable to the magnetization parameter of the free wind from the magnetar. Our estimate for $\sigw\sim 7\,\eta_2^{2/3}$ and $\eta\sim 10^2-10^4$ (\Sect~\ref{wind}) then implies that $\E_O\sim 10^{44}$~erg is possible for strong explosions, if they happen to strike the wind bubble in the tail of a previous flare. This optical flash is emitted on the characteristic timescale comparable to 1~s (\Eq~\ref{eq:nus3}).

%################################################

\section{Radio burst}
\label{maser}

\subsection{Shock transition and maser instability}

The wind magnetic field is transverse to the radial direction and parallel to the shock plane of the blast wave. The shock is collisionless and mediated by Larmor rotation, so its thickness is comparable to the gyro-radius. Such shocks are capable of emitting semi-coherent electromagnetic waves at the Larmor frequency before the plasma is thermalized into the downstream Maxwellian distribution \citep{Langdon88,Gallant92,Iwamoto17,Plotnikov19}. 

The jump conditions for strongly magnetized shocks ($\sigw\gg 1$) imply that the shock runs with Lorentz factor $\sigw^{1/2}$ relative to the downstream/postshock plasma (e.g. \citealt{Gallant92}). In the fixed lab frame, the shock Lorentz factor is 
\beq
\label{eq:Gsh}
   \Gsh\approx 2\Gamma\sigw^{1/2} \qquad (\sigw\gg 1),
\eeq
where $\Gamma$ is the Lorentz factor of downstream plasma (the blast).

When viewed in the blast frame, the upstream plasma (pre-shock wind) forms a cold ultra-relativistic beam. It moves with the drift speed of the electromagnetic field $\bv_D/c=\bE\times\bB/B^2$ and has no pre-shock gyration motion. Its Lorentz factor relative to the blast is
\beq
  \Grel \approx \frac{\Gamma}{2\Gw} \qquad (\Gamma\gg\Gw\gg 1).
\eeq
At the shock, the drift speed of the magnetic field lines joining the blast drops to zero, and the cold beam starts to gyrate with Lorentz factor $\Grel$ and Larmor frequency  
\beq
\label{eq:omL}
  \omL=\frac{c}{\rL}=\frac{e\B}{\Grel m_ec}  \quad {\rm (blast~frame)}.
\eeq
Here 
\beq
\label{eq:Bb}
   \B\approx 2\Grel\tBw    
\eeq
is the magnetic field in the blast (measured in the downstream/blast frame) found from the shock jump conditions at $\sigw\gg1$, and $\tBw$ is the pre-shock magnetic field of the wind (measured in its rest frame). \Eqs~(\ref{eq:omL}) and (\ref{eq:Bb}) imply that $\omL$ in the blast is related to the gyro-frequency in the pre-shock wind by 
\beq 
  \omL=2\tilde{\omega}_c=2\,\frac{e\tilde{B}_w}{m_ec}.
\eeq

The cold $e^\pm$ beam brought by the upstream and gyrating at the shock transition forms a ring-like structure in the momentum space. This ring is unstable to bunching \citep{Hoshino91,Gallant92}. The instability growth rate is a substantial fraction of the Larmor frequency, and the ring is quickly destroyed behind the shock, leading to complete thermalization of the plasma. As the shock moves ahead, the ring is continually reformed at the new shock location and continually destroyed behind it, leaving behind the hot Maxwellian $e^\pm$ plasma.  

The destruction of the $e^\pm$ ring occurs through emission of low-frequency ($\sim\omL$) electromagnetic waves by the unstably growing bunches  (e.g. \cite{Hoshino91}). This maser instability first converts the free energy of the ring to low-frequency waves, and then most of the waves get re-absorbed by the plasma and thermalized. However, some waves escape into the cold upstream and then to a distant observer.

This maser emission is efficient only if the upstream temperature is not relativistic; otherwise the thermal dispersion broadens the ring, and its bunching and emission of large-amplitude waves is suppressed (\cite{Amato06}; A. Babul and L. Sironi, in preparation). Therefore, as in Paper~I,  we focus on blast waves in the freely expanding, cold wind. The maser mechanism becomes inefficient if the blast wave hits the tail of a previous flare --- then it enters the hot wind bubble and turns into a bright source of synchrotron radiation described in \Sect~\ref{optical}.

\subsection{Doppler effect}
\label{Doppler}

The strong electromagnetic waves emerging ahead of the shock have a chance to escape while waves behind it (inside the thermalizing blast) are mostly destroyed \citep{Gallant92}. Since the shock is moving with Lorentz factor $\sigw^{1/2}$ with respect to the blast, only waves emitted at angles $0<\sin\theta<\sigw^{-1/2}$ (measured in the blast frame) are capable of overtaking the shock and escaping into the upstream. The limited range of emission angles implies a reduced range of Doppler factors for the frequency transformation from the blast frame to the lab frame,
\beq
\label{eq:D}
  D=\Gamma(1+\beta\cos\theta),  \quad 
  2\Gamma\left(1-\frac{1}{4\sigw}\right) <D< 2\Gamma.
\eeq
The strong magnetization of the wind $\sigw\gg 1$ implies $D\approx 2\Gamma$.
If the shock maser has a narrow spectral feature in the blast frame, its Doppler transformation to the observer frame will not smear out the feature.

The corresponding range of emission angles in the lab frame is given by 
\beq
  c\cos\theta>\vsh \qquad {\rm  or} \quad  \sin\theta<\Gsh^{-1}, 
\eeq
where $\vsh/c=(1-1/2\Gsh^2)^{1/2}$ is the shock speed. This implies that a distant observer can see the upstream maser waves from a small fraction of the (spherically curved) shock. This part of the sphere is almost exactly perpendicular to the line of sight, within angle $\thb$, and has area $r^2\Omb$, where 
\beq
  \Omb=2\pi(1-\cos\thb)\approx \frac{\pi}{\Gsh^2}.
\eeq
The arrival time of the waves, $\tobs$, is related to emission radius $r$ by
\beq
\label{eq:dtobs}
  d\tobs\approx \left(1-\frac{\vsh\cos\theta}{c}\right)\frac{dr}{\vsh} \approx \frac{dr}{\Gsh^2 c}\approx \frac{\Omb}{\pi}\,\frac{dr}{c}.
\eeq
Note that $\Omb$ also equals the solid angle occupied by the electromagnetic wave beam immediately outside the shock.

\subsection{Peak frequency}

Recently, simulations of the shock maser in $e^\pm$ plasma have been performed for moderately magnetized plasma with $\sigma < 1$ \citep{Iwamoto17,Iwamoto18}, and high magnetizations $\sigma$ up to 30 \citep{Plotnikov19}. It was found that the spectra of high-$\sigma$ shock masers have rather pronounced peaks, and the peak angular frequencies are higher than $\tilde{\omega}_c$ by a factor of $\xi\sim$ a few. In approximate agreement with \cite{Gallant92}, \cite{Plotnikov19} find $\xi\sim 3$ at high $\sigma$.

The corresponding peak frequency in the fixed lab frame (and the frame of a distant observer) is Doppler shifted by the blast wave motion with Lorentz factor $\Gamma$. The Doppler shift is $2\Gamma$ (\Eq~\ref{eq:D}). This gives the peak temporal frequency of observed waves,
\beq
\label{eq:nupk}
   \nupk=\xi\,\frac{\Gamma\tilde{\omega}_c}{\pi}=\frac{\xi}{\pi}\,\frac{e}{m_ec}
   \left(\frac{\Lw}{c r^2}\right)^{1/2} \frac{\Gamma}{\Gw}.
\eeq
It is a factor of $2\xi$ higher than the estimate in Paper~I, $\nupeak\sim \Gamma\omL/2\pi$, which neglected the factor of $\xi$ for the dominant emitted harmonic and the additional factor of 2 in  the Doppler shift of the beamed wave. 

The evolution of $\Gamma(r)$ is given by  
(\Sect~\ref{bw}),
\begin{eqnarray}
\label{eq:G3}
\Gamma(r)\approx \Gdec
\times \left\{\begin{array}{ll}
             1     & \quad r<\Rdec \\ 
   (\Rdec/r)^{1/2} & \quad r>\Rdec \\
                                                                                         \end{array}\right.
\label{eq:G5}
\end{eqnarray}
\beq
\label{eq:Gdec}
   \frac{\Gdec}{\Gw}= \Gamma_0=\left(\frac{\E}{\tau\Lw}\right)^{1/4} =10^2\left(\frac{\E_{44}}{\tau_{-3}L_{\rm w,39}}\right)^{1/4}.
\eeq
Substitution of $\Gamma(r)$ into \Eq~(\ref{eq:nupk}) gives $\nupk(r)$. In particular, at radii $r>\Rdec$ we find
\beq
   \nupk \sim \frac{e\,\Gw\,(2\E)^{1/2}}{m_ec\, r^{3/2}}
       \approx  \frac{5{\rm ~GHz}}{r_{14}^{3/2}}\,\left(\frac{\Gw}{20}\right)
       \E_{44}^{1/2}.
\eeq
$\Lw$ drops out from this relation.

From an observational point of view it is useful to express the $\nupk$ evolution in terms of arrival time $\tobs$, which is compressed by the Doppler effect. It is related to emission  radius $r$ by
\begin{eqnarray}
   \tobs(r)\sim \frac{r}{c\Gsh^2}=\tdec \times \left\{\begin{array}{ll}
         r/\Rdec     & \quad r<\Rdec \\
      (r/\Rdec)^2  & \quad r>\Rdec \\
                                                                                          \end{array}\right.
\label{eq:tobs2}
\end{eqnarray}
\beq
   \tdec\sim \frac{\tau}{2\sigw} =\frac{1 \rm ~ms}{2\sigw}\,\tau_{-3}.
\eeq
(Note that the arrival time is shorter than estimated in Paper~I by the factor of $\sim (2\sigw)^{-1}$.)

Then we find
\begin{eqnarray}
 \nupk= \nudec \times \left\{\begin{array}{ll}
           \tdec/\tobs              &  \quad \tobs<\tdec \\
           (\tdec/\tobs)^{3/4}  &  \quad \tobs>\tdec 
                                        \end{array}\right.
\label{eq:nupk1}
\end{eqnarray}
\begin{eqnarray}
\nonumber
 \nudec &\sim& \frac{e\,\Lw^{3/4}}{2 m_ec^{5/2}\, \E^{1/4}\,\tau^{3/4}\,\Gw^2} \\
       &\approx& 1.4\, \frac{L_{\rm w,39}^{3/4}}{\E_{44}^{1/4}\tau_{-3}^{3/4}}\left(\frac{\Gw} {20}\right)^{-2} {\rm GHz}.
      \label{eq:nudec}
\end{eqnarray}

\subsection{Luminosity and energy spectrum}
\label{luminosity}

As the blast wave propagates distance $dr$ in the wind with particle outflow rate $\dN$, it sweeps up $dN$ particles,
\beq
  dN\approx \frac{\dN\,dr}{2\Gw^2 c},
\eeq
and dissipates energy $d\E_{\rm diss}=2\Gamma\,(\Grel-1)  m_ec^2\,dN$ (measured in the lab frame). 
The upstream waves carry away a fraction $\eff$ of the dissipated energy,
\beq
\label{eq:dE}
  d\EFRB=\eff\, 2\Gamma\, (\Grel-1)  m_ec^2\,dN\approx\frac{\eff\,\Gamma^2 \Lw\,dr}{2\Gw^4\sigw c}.
\eeq
\cite{Plotnikov19} measured $\eff$ in their simulations and found $\eff\sim 10^{-3}/\sigma$ at high magnetizations $\sigma\gg 1$; numerical estimates below will be based on this formula.
A simple interpretation of the measured scaling $\eff\propto\sigma^{-1}$ is the reduced solid angle $\sim \sigma^{-1}$ of the ``escape cone'' for the waves (\Sect~6.2). 

In observer time, the blast wave propagation $dr$ takes $d\tobs\sim dr/\Gsh^2 c$, and the observed maser luminosity is
\beq
\label{eq:L}
   \LFRB=\frac{d\EFRB}{d\tobs}\sim 2\eff\,\frac{\Gamma^4}{\Gw^4}\,\Lw.
\eeq
Expressing $\LFRB$ as a function of $\tobs$, we find
\begin{eqnarray}
 \LFRB \sim \Ldec \times \left\{\begin{array}{ll}
          1                 &  \quad \tobs<\tdec \\
           \tdec/\tobs  &  \quad \tobs>\tdec 
                                        \end{array}\right.
\label{eq:L1}
\end{eqnarray}
\beq
\label{eq:Ldec}
  \Ldec\sim \eff\,\frac{\E}{\tau}\sim 10^{44}\,\frac{\E_{44}}{\sigw\,\tau_{-3}} \;\frac{\rm erg}{\rm s}.
\eeq

Distribution of the total emitted energy $\EFRB$ over emission frequency may be found from \Eq~(\ref{eq:dE}), taking into account the relation between $r$ and $\nupk$.
This gives
\begin{eqnarray}
   \frac{d\EFRB}{d\ln\nu} \sim \Edec \times \left\{\begin{array}{ll}
          \nudec/\nu               &  \quad \nu>\nudec \\
         1  &  \quad \nu<\nudec 
                                        \end{array}\right.
\label{eq:Enu}
\end{eqnarray}
\beq
    \Edec\sim \Ldec\tdec\sim \frac{\eff\,\E}{\sigw}\sim 10^{41}\,\sigw^{-2}\,\E_{44} {\rm ~erg}.
\eeq
When this time-integrated spectrum is resolved in time, one should observe the drift of $\nupk(\tobs)$ decreasing with $\tobs$ according to \Eq~(\ref{eq:nupk1}).

\subsection{Polarization}
\label{sec:polarization}

The shock maser was found to produce linearly polarized waves with the electric field perpendicular to the wavevector $\bk$ and the magnetic field $\bB$ (the ``extraordinary'' polarization mode). In 1D simulations, $\bk$ is constrained to be aligned with the shock propagation direction, and the waves have the 100\% linear polarization. Multi-dimensional simulations are required for more reliable measurement of the angular distribution and the polarization degree of the escaping waves. \cite{Gallant92} reported 2D results for one value of magnetization $\sigma$ and with magnetic fields perpendicular to the simulation plane. They found that the escaping waves remain strongly beamed along the shock direction in the 2D simulation, and that the waves have a 97\% polarization in the extraordinary mode. \cite{Iwamoto18} extended 2D simulations to configurations with magnetic field in the simulation plane and found that both linear modes are generated. Full 3D simulations would help verify whether the shock maser is capable of producing waves with extremely high polarization dominated by the extraordinary mode.
 
The magnetic field in the wind from the rotating magnetar is wound up into a tight spiral around $\boldsymbol{\Omega}$. Far outside the light cylinder the field direction is almost exactly toroidal, $\bB=B\boldsymbol{e}_\phi$, and so the extraordinary wave polarization vector $\boldsymbol{k}\times\bB$ is perpendicular to $\boldsymbol{e}_\phi$, i.e. is in the plane defined by the line of sight and $\boldsymbol{\Omega}$.
Therefore, the observed polarization vector of the extraordinary mode is aligned with the magnetar rotation axis projected onto the plane of the sky.

\subsection{Temporal variations}
\label{sec:temporal}

Observed FRBs last up to milliseconds and show variable light curves \citep{Hessels19}. Below we briefly discuss variability that may be expected from the blast wave in a wind.

Note that the pre-explosion wind forming the ambient medium for the blast wave differs from the persistent wind (\Sect~2) and may be unsteady. Variations in $\Lw$ or $\Gw$ will create inhomogeneities in magnetic pressure, which can excite magnetosonic waves. The waves propagate with Lorentz factor $\sigw^{1/2}$ through the cold magnetized wind, and thus redistribute the magnetic pressure with nearly speed of light. Causal contact (in the outward direction) across a wind shell of thickness $\delta r$ in the lab frame takes time $t\sim \delta r/2\Gw^2c $. A modulation of the wind on a scale $\delta r$ enters the causal horizon and ``unfreezes,'' becoming a propagating magnetosonic wave, at radius $r\sim ct\sim 2\Gw^2\delta r$. 

Thus, at a given radius $r$, variations in $\Lw$ and $\Gw$ persist on scales as small as $\delta r\sim r/2\Gw^2$. Note that the blast wave sweeps up a shell of similar thickness $r/2\Gw^2$ as it expands to radius $r$. Thus, pre-explosion wind variations on this scale will not have a chance to fully relax before becoming swept up by the (super-magnetosonic) shock. Magnetosonic waves will have enough time to smoothen the radial profiles of $\Gw$ and $\Lw$ on scales $\delta r\ll r/2\Gw^2$. However, the relaxation of pressure waves on the short scales still does not erase variations in $\sigw$; in fact it may amplify them.

We conclude that, at a given time $t=r/c$, the blast wave sweeping a variable wind may encounter (1) strong but smooth variations in $\Lw$ and $\Gw$ on timescales comparable to $t$, and (2) strong short-timescale variations in $\sigw$. This will lead to variable FRB emission. The arrival time of waves emitted at radius $r$ (not taking into account any additional propagation effects outside the source) is given by 
\beq
   \tobs(r)\sim \int_0^r\frac{dr^\prime}{c\,\Gsh^2(r^\prime)}
    \approx  \frac{1}{c\,\Gamma^2(r)} \int_0^r\frac{dr^\prime}{\sigw(r^\prime)},
\eeq
where we took into account that $\Gamma(r)$ is a smoothly varying function (and so may be taken out of the integral) while $\sigw$ may have large variations on short scales. The integral depends on the distribution of $\sigw$ ahead of the blast wave. If $\sigw$ has a broad distribution, the observed radiation tends to be dominated by the parts with lowest $\sigw$, because they have the highest radiative efficiency. These parts also give the longest duration $\delta\tobs\propto\sigw^{-1}$. For instance, if half of shell $\delta r=r/2\Gw^2$ has $\sigw=\sigma_1$ and half has $\sigma_2>\sigma_1$, the observed radiation will be dominated by the $\sigma_1$ part and will have the duration 
\beq
  \tobs\sim \frac{1}{2}\,\frac{r}{\Gamma^2\sigma_1}.
\eeq

\subsection{Strength parameter of the maser waves}

A particle exposed to the high-amplitude electromagnetic waves ahead of the shock will oscillate with a high momentum. The strength of its acceleration is described by the dimensionless parameter, 
\beq
 a=\frac{eE}{m_ec\,\omega},
\eeq
where $E$ is the amplitude of the wave electric field; $\bE$ is perpendicular to the wavevector $\bk$, which is nearly radial. The strength parameter is invariant under Lorentz boosts along $\bk$, i.e. along the radial direction, and therefore it is the same in the blast frame, wind frame, and lab frame. The linear description of the waves (Fourier modes with wavevectors $\bk$) is approximately valid as long as $a<1$.

The wave amplitude $E$ is related to its average energy flux $F=E^2/8\pi$ and may be expressed in terms of the FRB luminosity $\LFRB$. This relation (in the lab frame) is 
\beq
   E=\left(\frac{2\LFRB}{cr^2}\right)^{1/2}.
\eeq
Substituting \Eq~(\ref{eq:L}) for $\LFRB$, and using the maser peak frequency $\omega=2\Gamma\xi \tilde{\omega}_c$ in the lab frame, we find
\beq
   a=\frac{\eff^{1/2}}{\xi}\,\frac{\Gamma}{\Gw},
\eeq
where $\xi\sim 3$, and $\Gamma(r)$ is given in \Eq~(\ref{eq:G3}). This gives
\begin{eqnarray}
\label{eq:a}
a(r)\approx \adec
\times \left\{\begin{array}{ll}
             1     & \quad r<\Rdec \\
             (\Rdec/r)^{1/2} & \quad r>\Rdec \\
                                                                                         \end{array}\right.
\end{eqnarray}
\beq
   \adec= \frac{\eff^{1/2}}{\xi}\,\Gamma_0 \sim 1\;\eff_{-3}^{1/2} \left(\frac{\E_{44}}{\tau_{-3}L_{\rm w,39}}\right)^{1/4}.
\eeq
The characteristic $a\simlt 1$ means that the emitted FRB wave propagating through the wind forces the cold wind plasma to oscillate with mildly relativistic speeds.

%###########################################################

\section{Induced Compton scattering}
\label{ind}

Spontaneous Thomson scattering of electromagnetic waves in the wind is very weak. Thomson optical depth $\tauT(r)$ seen by a beam emitted at a radius $r$ is a Lorentz invariant quantity, which may be evaluated in any frame. Travelled path $dr$ in the lab frame corresponds to path $cd\tilde{t}=\Gw(1-\beta_{\rm w})dr$ in the wind frame, density $n_\pm$ corresponds to $\tilde{n}_\pm=n_\pm/\Gw$ in the wind frame, and
\beq
  d\tauT=\sT \tilde{n}_\pm cd\tilde{t}= (1-\beta_{\rm w})\sT n_\pm dr,
\eeq
where $\sT=(8\pi/3)r_e^2$ is the Thomson cross section. Substituting $n_\pm=\dN/4\pi r^2 c$ and integrating over $dr$ from the emission radius $r$ to infinity one finds 
\beq
\label{eq:tauT}
   \tauT(r)\approx \frac{\sT \dN}{8\pi c r \Gw^2}
   \approx 10^{-10}\, \frac{\dN_{42}}{r_{14}\,\Gamma_{\rm w,1}^2}.
\eeq

Raman scattering is possible in the presence of ions. However, the ion component of the spindown magnetar wind (estimated \Eq~\ref{eq:ions}) is tiny compared with the $e^\pm$ component. Therefore, Raman scattering is negligible.

Induced Compton scattering can be much stronger and presents a threat for low-frequency waves, as it tends to damp the beam \citep{Zeldovich69,Blandford75,Wilson78}.  Induced scattering can occurs in two ways. (1) Scattering inside the beam: both initial and final states of the photon have propagation directions within the beam, i.e. very close to the radial direction. (2) Scattering outside the beam: the final state has arbitrary propagation direction; its occupation number is seeded by spontaneous Thomson scattering and then exponentially amplified by induced scattering \citep{Coppi93,Lyubarsky08}. Below we estimate both effects.

\subsection{Induced scattering inside the beam}

Induced scattering removes energy from the FRB beam by shifting the wave spectrum to lower frequencies. The magnitude of this effect can be derived from the equation of time-dependent  radiative transfer applied to the blast wave radiation propagating through the wind. The derivation will be given elsewhere, and here we state the results in a simple intuitive form.

The frequency shift of the waves $\Delta\nu/\nu$ due to induced scattering  
scales with the plasma column density crossed by the beam, or its Thomson optical depth $\tauT$. Both $\Delta\nu/\nu$ and $\tauT$
are frame-independent quantities. The coefficient of proportionality between $\Delta\nu/\nu$ and $\tauT$ takes the simplest form in the rest frame of the scattering plasma, i.e. in the wind frame (where quantities are denoted with tilde),
\beq
\label{eq:ind0}
   \frac{\Delta\nu}{\nu}=-\alpha\, \N \frac{2h\nucom}{m_ec^2}(1-\cos\tthb)\, \tOmb\, \tauT
   \quad ({\rm inside~beam}).
\eeq 
Here $\N$ is the photon occupation number in the beam, $\tthb\ll 1$ is its opening angle, and $\tOmb$ is the solid angle occupied by the beam. Evaluating the numerical factor $\alpha$ involves detailed transfer calculations, which give $\alpha\sim 3\times 10^{-2}$.

The shift $\Delta\nu/\nu$ written in this form has a straightforward physical meaning, which was discussed previously in detail (e.g. \cite{Wilson78}). 
Induced scattering inside the beam involves an initial photon state A and a final state B that are both inside the solid angle 
\beq
  \tOmb=2\pi(1-\cos\tthb)\approx\pi\tthb^2.
\eeq 
Compared with spontaneous Thomson scattering (which would give the optical depth $\tauT$),  induced scattering A$\rightarrow$B is boosted by the occupation number in state B, $\N_B\gg 1$. The opposite induced scatterings A$\leftrightarrow$B occur with equal rates proportional to $\N_A\N_B$. If scattering was exactly coherent, the net effect of transitions A$\leftrightarrow$B would vanish. In fact, a small energy loss does occur in each scattering cycle A$\rightarrow$B$\rightarrow$A$^\prime$ as a result of electron recoil, which makes the photons drift downward in energy. For the typical angle between the initial and final photon states $\tth_{\rm AB}\sim\tthb$, the recoil effect is $\delta\nu/\nu=-(2h\nucom/m_ec^2)(1-\cos\tthb)$, which explains the appearance of this combination in  \Eq~(\ref{eq:ind0}). Finally, the factor of $\tOmb$ in \Eq~(\ref{eq:ind0}) results from integration over all possible scattering angles within the beam to obtain the net effect on the occupation number $\N=\N_A$ in a given state A. 

This intuitive description invokes quantum mechanics and the Planck constant $h$, however in the end the effect is purely classical: $h$ enters twice, in the recoil effect and the definition of  occupation number, and drops out in the final result. The effect is more concisely expressed in terms of the dimensionless brightness temperature $\tThB=\N h\nucom/m_ec^2$,
\beq
\label{eq:ind1}
   \frac{\Delta\nu}{\nu}=-\frac{\alpha}{\pi}\, \tThB \tOmb^2\, \tauT.
\eeq 
The brightness temperature of waves in a single polarization state with intensity $I_\nu$ in the lab frame is
\beq
\label{eq:ThB}
  \ThB\equiv \frac{kT_B}{m_ec^2}=\frac{I_\nu}{m_e\nu^2}.
\eeq
Using $kT_B=\N h\nu$, where the photon occupation number $\N$ is a Lorentz-invariant quantity, one can see  that $\ThB$ transforms under Lorentz boosts as $\nu$, 
\beq
  \tThB=\frac{\nucom}{\nu}\,\ThB\approx \frac{\ThB}{2\Gw}.
\eeq
The solid angles occupied by the beam in the lab and wind frames are related by $\tOmb=(\nu/\nucom)^2\,\Omb\approx 4\Gw^2\Omb$. Using these transformations, one can express $\Delta\nu/\nu$ in terms of the lab frame quantities,
\beq
\label{eq:ind2}
   \frac{\Delta\nu}{\nu}=-\frac{8\alpha}{\pi}\,\Gw^3 \ThB \Omb^2\, \tauT.
\eeq 

It remains to evaluate the brightness temperature of the waves emitted by the blast wave. The wave energy emerges ahead of the shock (at $\cos\theta>\beta_{\rm sh}$) with rate 
\beq
\label{eq:Inu}
  \frac{c}{4\pi r^2} \frac{d\E_\nu}{dr}=2\pi \int_{\beta_{\rm sh}}^{1} I_\nu(\theta)(\cos\theta-\beta_{\rm sh})\, d\cos\theta
  \approx \frac{I_\nu\Omb^2}{4\pi}.
\eeq
The spectrum of shock maser emission is convenient to write in the form
\beq
  \nu\frac{d\E_\nu}{dr}=f(\nu)\,\frac{d\EFRB}{dr}, \qquad \int f(\nu)\,d\nu=1.
\eeq
Then from \Eqs(\ref{eq:ThB}) and (\ref{eq:Inu}) one finds
\beq
\label{eq:ThOm}
   \ThB(\nu)\,\Omb^2\approx \frac{c\,f(\nu)}{m_e\nu^3r^2}\,\frac{d\EFRB}{dr}.
\eeq
The rate of wave energy emission $d\EFRB/dr$ is determined by the efficiency $\eff\sim 10^{-3}\sigw^{-1}$ as discussed in \Sect~\ref{luminosity},
\beq
\label{eq:dEdr}
  \frac{d\EFRB}{dr}=\eff\,\frac{\Gamma^2}{2\Gw^3}\,m_ec\,\dN.
\eeq

Using \Eq~(\ref{eq:tauT}) for $\tauT$ and \Eq~(\ref{eq:ThOm}) for $\ThB\Omb^2$ we find from \Eq~(\ref{eq:ind2}),
\beq
  \label{eq:ind3}
   \frac{\Delta\nu}{\nu}\approx -\frac{\alpha}{2\pi^2}\,\eff\,\frac{c\sT}{r^3\nu^3}\,f(\nu)\,\dot{N}^2\,\frac{\Gamma^2}{\Gw^2}.
\eeq
In particular, at the spectral peak, using $f(\nupk)\sim 1$ and $\nupk$ from \Eq~(\ref{eq:nupk}), we obtain the final result
\beq
 \frac{\Delta\nupk}{\nupk}\sim -10^{-2}\,\eff\,\frac{e}{m_ec^{5/2}}\,\frac{\Lw^{1/2}}{\eta^2}\,\frac{\Gw}{\Gamma},
\eeq
where we have used $m_ec^2\dN=\Lw/\eta$ and $\sT=(8\pi/3)(e^4/m_e^2c^4)$.

Induced scattering does not strongly affect the FRB beam as long as $|\Delta\nupk/\nupk|<1$, which requires
\beq
  \eta>0.1\,\left(\eff\,\frac{e}{m_ec^{5/2}}\,\frac{\Gw}{\Gamma}\right)^{1/2} \Lw^{1/4}.
\eeq
In the blast waves from magnetars the ratio  $(\Gamma/\Gw)^{1/2}\sim 10$ is constant at $r<\Rdec$ and decreases as $(r/\Rdec)^{-1/4}$ at $r>\Rdec$ (\Eq~\ref{eq:G3}). Therefore, the constraint on $\eta$ rather weakly depends on the maser emission radius. Substituting $\Gamma/\Gw\approx\Gamma_0=(\E/\tau\Lw)^{1/4}$ and $\eff\approx 10^{-3}\sigw^{-1}\approx 3\times 10^{-3}\eta^{-2/3}$, we find
\beq
\label{eq:etacond}
  \eta>50\,L_{\rm w,39}^{9/32}\left(\frac{\E_{44}}{\tau_{-3}}\right)^{-3/32}.
\eeq
We conclude that blast waves in winds with $\eta>10^2$ will produce FRBs capable of escaping without significant induced downscattering inside the beam.

\subsection{Induced scattering outside the beam}

In addition to Bose condensation inside the beam, a similar induced process can exponentially amplify any seed radiation outside the beam \citep{Coppi93, Lyubarsky08}. Compared with induced scattering inside the beam, this process is accelerated by the large recoil factor $\Delta\nucom/\nucom\propto (1-\cos\tilde{\theta})$, as the scattering angle $\tilde{\theta}$ is no longer required to stay inside the narrow beam. As a result, induced scattering outside the beam may become dominant even though the density of seed radiation is much smaller than the beam density. A detailed calculation, which will be presented elsewhere, confirms the following simple estimate. 

Consider radially propagating waves with duration $\tobs$. When viewed in the wind rest frame, the waves fill a shell of thickness $\tilde{\varpi}\approx 2\Gw c\tobs$. Induced scattering exponentially amplifies seed waves outside the beam by the factor $\exp(r_e^2\tilde{\varpi}\tilde{n}_\pm \tilde{g^\prime}/m_e)$, where $\tilde{g}^\prime =(\nu/\tilde{\nu})g^\prime$ and 
\beq
\label{eq:gp}
 g^\prime\equiv\frac{\partial}{\partial \nu} \frac{F_\nu}{\nu}
   \sim \frac{\LFRB}{4\pi r^2\nu^2}. 
\eeq
The seed radiation is provided by spontaneous Thomson scattering of the primary beamed waves, and so it is weaker by the factor of $\sim\tauT$ compared with the beam. Induced scattering outside the beam does not significantly damp the beam if 
\beq
  \tauT \exp\left(\frac{r_e^2\tilde{\varpi}\tilde{n}_\pm \tilde{g^\prime}}{m_e}\right)<1.
\eeq
Substituting here $\tilde{\varpi}\approx 2\Gw c\tobs$, $\tilde{n}_\pm=\dN/4\pi r^2\Gw c$, and $\tilde{g}^\prime=2\Gw g^\prime$, one can rewrite this condition as
\beq
\label{eq:ind_outside}
    \frac{4r_e^2 \Gw \dN \tobs g^\prime}{\pi m_e r^2}<-\ln\tauT,
\eeq
Note that this condition can be violated only in the spectral region where $g^\prime(\nu)>0$. Therefore, the induced scattering outside the beam becomes particularly interesting if the FRB spectrum is observed to have a region where $g^\prime>0$.  

The condition~(\ref{eq:ind_outside}) gives an upper limit on the burst energy emitted at radius $r$,
\beq
\label{eq:Eind}
  \E_{\rm ind}\sim \frac{4\pi ^2 m_e\, |\ln\tauT|\, r^4 \nu^3}{r_e^2 \Gw\dN}.
\eeq
Two aspects of this limit should be noted. 
\\
(1) If the emitted FRB spectrum has $g^\prime(\nuobs)<0$ in the observed frequency window, the condition~(\ref{eq:ind_outside}) is initially not violated at $\nuobs$ even if $\EFRB>\E_{\rm ind}$. However, at some lower frequency $\nu_+$ the emitted spectrum is expected to break so that $g^\prime(\nu)>0$ at $\nu<\nu_+$. If $\EFRB(\nuobs)>\E_{\rm ind}(\nuobs)$, the condition~(\ref{eq:ind_outside}) will be inevitably violated at $\nu<\nu_+$. This  can be seen from the fact that at $\nu_+$ the ratio of $d\EFRB/d\ln\nu\propto \nu^2 g(\nu)$ to $\E_{\rm ind}(\nu)\propto \nu^3$  is larger than the same ratio at $\nuobs$. Then the question arises whether strong induced scattering at $\nu\simlt\nu_+$ drives spectral distortions toward higher frequencies by developing very high $g^\prime$ (with oscillations in sign), and further investigation requires detailed calculations including thermal Doppler effects due to non-zero plasma temperature. If such nonlinear evolution does occur at frequencies $\nu>\nu_+$, its likely outcome is the reduction of the burst energy to $d\EFRB/d\ln\nu\sim \E_{\rm ind}(\nu)$. Then the limit in \Eq~(\ref{eq:Eind}) may be effectively applicable at $\nuobs$ regardless of $g^\prime(\nuobs)$ in the originally emitted spectrum. 
\\
(2) The above estimates for induced scattering are safe where the waves have the strength parameter $a<1$. In this regime one may approximate the ambient plasma as a static medium in the appropriately chosen frame (the rest frame of the wind). In the opposite regime, $a>1$, the wave forces relativistic motions of the plasma on the wavelength scale, which makes induced scattering calculations problematic. The strength parameter is related to the burst luminosity $\LFRB$, frequency $\nu$, and radius $r$,
\beq
\label{eq:aFRB}
   a=\left(\frac{r_e\,\LFRB}{2\pi^2 c m_e r^2\nu^2}\right)^{1/2}
     \approx 0.6 \,\frac{L_{\rm FRB,43}^{1/2}}{r_{14}\,\nu_9}.
\eeq
One can see that the induced scattering estimates are typically reliable at radii $r\simgt 10^{14}$~cm, which is comparable to the FRB emission radius in the proposed blast-wave model (\Sect~\ref{maser}). 

In the domain of its applicability, the limit $\EFRB\simlt\E_{\rm ind}$ (\Eq~\ref{eq:Eind}) can be used to constrain the emission radius of a burst with a given observed energy $\EFRB$, 
\beq
\label{eq:Rind}
   r>R_{\rm ind}\sim 10^{14}\,\nu_9^{-3/4}\E_{\rm FRB,40}^{1/4}
   \left(\frac{\Gw}{20}\right)^{1/4}\dN_{42}^{1/4} {\rm cm}.
\eeq

%################################################################

\section{Discussion}

Giant flares are produced by sudden dissipation events in the magnetar magnetosphere, which develop high (mildly relativistic) temperatures. Therefore they are normally observed in the gamma-ray and X-ray bands.  The limited capabilities of existing gamma-ray detectors allow one to observe the flares only in the local universe, in our and nearby galaxies. This paper, however, shows  that the magnetospheric flares launch powerful blast waves in the magnetar wind capable of producing ultra-strong radio waves and bright optical radiation. This emission is detectable from large cosmological distances by radio and optical telescopes, because low-frequency instruments are much more sensitive than gamma-ray detectors. The low-frequency detections likely require a preferential orientation, as the blast waves are anisotropic and emit radiation within a limited solid angle. 

Particularly promising sources are the recently born, hyper-active magnetars (HAMs), which should flare much more frequently than the older magnetars discovered so far in our Galaxy. They are expected to launch multiple blast waves and produce multiple radio bursts.

\subsection{Summary of the model}

The blast waves are launched into the relativistically expanding medium --- the  wind from the rotating magnetar, whose structure is summarized in Figure~\ref{fig:wind}. A suitable zone for a blast wave to produce an FRB is the cold, helical-$\bB$ wind at large latitudes. The dissipative equatorial striped wind is not suitable, because the maser mechanism fails in shocks propagating in hot media.

Our estimates show that magnetar winds have much stronger $e^\pm$ loading compared with ordinary pulsars.
This results in a moderate wind energy per particle rest-mass, $\eta\sim 10^2- 10^4$, and Lorentz factor $\Gw\sim 3\eta^{1/3}\sim 10\,\eta_2^{1/3}$. These wind parameters are important for the picture of explosion from a giant flare, as they control the blast wave Lorentz factor $\Gamma$ and deceleration radius $\Rdec$.

The explosion is driven by an ultra-relativistic magnetic plasmoid ejected at the beginning of a giant flare. The plasmoid experiences strong side expansion while preserving thickness $\Delta\sim 10^7$~cm and accelerating to a huge Lorentz factor $\Gamma_{\rm pl}\simgt 10^5$. It drives a blast wave with $\Gamma\simgt 10^3$ into the surrounding wind. We find that the blast wave emits a GHz burst at radii comparable to its deceleration radius $\Rdec\sim 10^{14}\,(\Gw/20)^2\Delta_7$~cm (see \Eq~(\ref{eq:nudec})).

The detailed analysis in this paper supports the proposal of Paper~I that FRBs may be emitted by blast waves in magnetar winds. Paper~I argued that the frequent repeaters (such as FRB~121102) are young, hyper-active magnetars, and that a similar blast-wave mechanism may also produce rarely repeating FRBs. The model predicts bursts with the following properties.
\\

(1) {\bf Rate.} If most of observed FRBs are generated by magnetars, their rate in the universe is set by the rate of magnetar flares. Soon after the FRB discovery, it was noticed that the two rates are in approximate agreement 
\citep{Popov13}. The giant flare rate was roughly estimated from the three flares observed so far in our Galaxy and the LMC, and it may need to be revised to include the putative hyper-active magnetars in distant galaxies.  
Accurate comparison of the flare rate with the observed FRB statistics is also complicated by a few other factors, including beaming of FRB emission, which can hide a large fraction of bursts. The contribution of frequent repeaters to the total rate is poorly known which further complicates the comparison. Note that the frequent bursts are weaker than FRBs detected from non-repeaters (or not identified as repeaters yet, because they repeat rarely). In view of this diversity, it may be useful to quantify the FRB population by its net power, ${\cal P}=\bar{\E}_{\rm FRB}\dot{\N}_{\rm FRB}$, where $\bar{\E}_{\rm FRB}$ is the average FRB energy (isotropic equivalent).  
\medskip

(2) {\bf Energy budget.} During the expected life-time of hyper-activity, $t\sim 10^9$~s, magnetars may release magnetic energy in frequent weak flares or rare strong flares, with a similar total energy budget up to $\sim 10^{50}$~erg.\footnote{The energy budget of magnetars is discussed in Paper~I. Their rotational energy is not capable of feeding the observed emission, and their magnetic energy can be sufficient. The observed magnetars likely have internal fields $B\sim 10^{16}$~G \citep{Kaspi17} which corresponds to the energy budget of $10^{49}$~erg. The young HAMs can have even stronger fields and thus more energy.} 
A significant fraction of this energy is released in magnetic plasmoids ejected explosively from the twisted magnetosphere. This ejection is sufficient to explain the energetics of observed FRBs, including frequent repeaters, with a realistic efficiency of radio emission. Note that typical bursts from FRB~121102 have $\EFRB\sim 10^{39}-10^{40}$~erg \citep{Law17}, which is a tiny fraction $\sim 10^{-10}-10^{-9}$ of the putative magnetar energy. 
\medskip

(3) {\bf Radiative efficiency.}
The observed $\EFRB$ is determined by the isotropic equivalent of the explosion energy $\E$ and the efficiency of radio emission. Our model predicts that the efficiency $\EFRB/\E$ is controlled by the wind magnetization parameter $\sigw=\eta/\Gw\sim 0.3\,\eta^{2/3}\sim 7\,\eta_2^{2/3}$. The theoretical FRB energy is then given by $\EFRB\sim 10^{-3}\,\sigw^{-2}\,\E$ (\Sect~\ref{maser}; Plotnikov \& Sironi 2019). Thus, efficiencies $\EFRB/\E\sim 10^{-5}-10^{-6}$ are expected, and the magnetar is capable of producing $N$ radio bursts with energies $\EFRB$ such that $N\EFRB\sim 10^{44}$~erg. For instance, $\sim 10^5$ bursts may be produced with energies $\EFRB\sim 10^{39}$~erg over the life-time of hyper-activity.
This number is consistent with observations of FRB~121102.
\medskip

(4) {\bf GHz frequencies.} The maser emission peaks at the frequency $\nupk$ that scales with the local magnetic field in the wind (\Eq~\ref{eq:nupk}). The  magnetic field decreases as the blast wave expands and therefore $\nupk$ sweeps from high to low frequencies. This prediction of the blast wave model was made in Paper~I and discussed in more detail in \Sect~6.3 (see also \cite{Metzger19}). At some radius $R_{\rm GHz}$, $\nupk$ passes through the GHz band. For typical parameters of explosions into magnetar winds, $R_{\rm GHz}$ is comparable to the blast wave deceleration radius $\Rdec$. By this time, most of the explosion energy is deposited into the blast wave, leading to the maximum radiative efficiency of GHz emission. However, we point out that  magnetar blast waves are capable of producing bursts with $\nupk$ far from 1~GHz.
\medskip

(5) {\bf Duration.} The GHz burst ends when the blast wave expands beyond $R_{\rm GHz}$, so that the emission frequency drops below the GHz band. The apparent duration of FRB emission is strongly compressed by the Doppler effect, $\delta \tobs\sim R_{\rm GHz}/c\Gsh^2$. Our estimates predict $\delta \tobs$ shorter than 1~ms, because we find an extremely high Lorentz factor of the shock, $\Gsh\simgt 10^4$. This suggests that we have overestimated $\Gsh$ by a factor of $\sim 3$ or that the typical observed duration $\delta\tobs\sim 1$~ms  is increased by propagation effects outside the source, apart from the standard dispersion effect. On the other hand, bursts as short as $30\,\mu$s have been detected \citep{Petroff19}. Furthermore, \cite{Connor19} recently argued that the typical $\sim 1$~ms may result from insufficient temporal and spectral resolution, and that many FRBs may have intrinsic durations much shorter than 1~ms.
\medskip

(6) {\bf Variability.} It is plausible that the  duration and temporal structure of FRBs are influenced by wave propagation through the ionized gas of the host galaxy, which results in lensing effects \citep{Cordes17}. Lensing could lead to a complicated temporal and spectral structure. However, intrinsic variatiability in the blast-wave emission is also possible, as discussed in \Sect~\ref{sec:temporal}. It is caused by variability in the pre-explosion wind ahead of the blast wave. Note also that the wind is modulated with the rotation period of the magnetar, $P\sim 1$~s. This could result in sub-ms periodicity in FRB emission, since the blast wave emission occurs with time compression by the Doppler effect by the factor of $\sim (\Gw/\Gsh)^2\sim 10^{-4}\sigw^{-1}$.
\medskip

(7) {\bf Induced scattering.}  As shown in \Sect~\ref{ind}, induced scattering does not suppress the FRB emission from the blast wave in the wind, as long as the wind has the energy parameter $\eta>10^2$. This condition should be satisfied by magnetar winds, which are estimated to have $\eta\sim 10^2-10^4$ (\Sect~\ref{wind}).
\medskip

(8) {\bf Spectrum.}
The time-integrated spectra of FRBs predicted by the model are described by \Eq~(\ref{eq:Enu}). The predicted spectral slope $d\ln \EFRB/d\ln\nu$ changes  from $-1$ to $-2$ at the characteristic  frequency $\nudec=\nupk(\Rdec)$, which is given by \Eq~(\ref{eq:nudec}). Similar slopes were reported in FRBs \citep{Macquart19}. However, narrow-band spectra with extremely steep slopes were also reported e.g. in FRB~121102 \citep{Hessels19}. A blast wave might produce such spectra only if its emission strongly peaks at some radii, at localized drops of $\sigw$ in the inhomogeneous wind. However, drawing conclusions may be premature, as it is unclear if the observed spectral features are intrinsic to the source or result from propagation effects. 
\medskip

(9) {\bf Polarization.}
The shock maser is expected to produce linearly polarized waves (\Sect~\ref{sec:polarization}). After the correction for propagation effects (in particular Faraday rotation), the polarization angle is expected to stay constant for all bursts in a repeater, because it is set by the direction of the magnetar angular velocity $\bOm$. This aspect is consistent with observations of FRB~121102 (Michilli et al. 2018). However, the model is hardly capable of explaining the diverse polarization data for FRBs \citep{Petroff19} without invoking more subtle propagation effects, such as conversion from linear to circular polarization.
\medskip

(10) {\bf Optical bursts.}
The optical and X-ray emission from blast waves in magnetar winds should be normally weak and hard to detect. However, in frequent repeaters, a dramatic enhancement of optical emission can happen when the blast wave strikes the wind bubble in the tail of a preceding flare (\Sect~\ref{optical}). Our estimates show that the optical luminosity in extreme cases can reach that of supernovae Ia, and last $\sim 1$~s. This result suggests that optical flashes should be looked for in frequent repeaters, such as FRB~121102. The optical flashes can be detected with future instruments searching for short optical transients. Their expected rate is a small fraction of the FRB rate. 
\medskip

(11) {\bf Location in host galaxies.} The observed local magnetars are associated with collapse of massive stars. They are found in regions of active star formation and often located inside a visible supernova remnant. This fact does not exclude formation of magnetars with no association with active star formation, as there are other plausible formation scenarios with old progenitors. The old progenitors have sufficient time to move away from their original location in the host galaxy. In particular, mergers of compact binaries --- two neutron stars, two white dwarfs, or a neutron star and a white dwarf --- may produce FRB-emitting magnetars. In addition, accretion-induced collapse of white dwarfs could create active magnetars. The possible contributions of compact progenitors to the FRB population are discussed in more detail in Margalit et al. (2019). 

Magnetars formed by compact mergers should be expected to display hyper-activity, because they have enormous differential rotation at birth, generating ultra-strong magnetic fields, and are more massive than canonical magnetars. Both factors accelerate the magnetic field evolution during the first decades of the magnetar life (Beloborodov \& Li 2016) and likely result in hyper-activity.
Magnetars formed through the canonical channel of stellar collapse may also produce unusually strong activity, in particular if the progenitor had a low metallicity (hence a weak stellar wind and a high retained angular momentum). 

The different magnetar formation channels might explain the observed diversity of FRB hosts. The dwarf host galaxy of FRB~121102 has a high star formation rate and a low metallicity \citep{Tendulkar17}. In contrast, three new localizations of FRB sources (not observed to repeat yet) are associated with older populations in massive galaxies \citep{Bannister19,Prochaska19,Ravi19}. Theoretical predictions for the FRB  rates from different populations are unfortunately uncertain. Magnetars with compact binary progenitors could be prolific bursters, however their low abundance in the universe could reduce their detection rate. It remains unclear which channel should dominate the formation of frequent repeaters. The only localized repeater, FRB~121102, appears consistent with a product of massive star collapse; however more localizations are needed to clarify the progenitors of frequent repeaters. 
\medskip

The non-detection of any radio signal from the powerful giant flare of SGR~1806-20 in the Milky Way \citep{Tendulkar16} demonstrates that not every giant flare produces an observable FRB.
Several factors can prevent detections of radio and optical flashes from local magnetars.
(1) Observations are sparse. Only three giant flares were seen so far, and for only one of them upper limits in the radio band are available.
(2) Ultra-relativistic blast waves from ejected plasmoids are strongly anisotropic, and their emission is Doppler beamed. The plasmoid needs to be ejected in our direction in order to see its emission. 
(3) A radio burst is expected if the blast wave propagates in the cold helical-$\bB$ zone of the wind, far from the equatorial plane of the rotating magnetar (Figure~\ref{fig:wind}). 
(4) Emission depends on the wind parameters. Winds from the old local magnetars differ from hype-active magnetar winds, in both power $\Lw$ and Lorentz factor $\Gw$.
(5) The bright optical flashes predicted in \Sect~\ref{optical} require two consecutive giant flares with a time separation of $\simlt 1$~day. This can happen in HAMs but is unlikely for the less active local magnetars.

\subsection{Comparison with Metzger et al. (2019)}

The recent paper by \cite{Metzger19} develops the FRB scenario of Paper~I with the focus on shocks between two subsequent giant flares, separated by $t_{\rm sep}\sim 10^5$~s. They consider a self-similar blast wave generated by the second flare and propagating in the magnetized ($\sigma\sim 1$), sub-relativistic (effectively stationary) ion tail of the first flare. The possibility of two-flare interactions was also suggested in Paper~I, and a more detailed analysis is presented in \Sects~\ref{bubble} and \ref{impact}. 

There are three issues, which make it hard to produce FRBs by blast waves in slow tails: 
\medskip

(1) The tail in fact does not sustain $\sigma\sim 1$, because its radial spreading reduces magnetization $\sigma$ to very low values (\Eq~\ref{eq:sigt}).  Then the shock maser mechanism should not operate for a blast wave propagating in the tail of a previous flare.
\medskip

(2) The picture of an explosion into a slow self-similar tail occupying radii $10^{11}<r<10^{15}$~cm \citep{Metzger19} gives a moderate Lorentz factor of the blast wave $\Gamma\sim 10^2$ (see \Eq~(26) in Paper~I), and then the short observed duration $\tobs\sim  r/\Gamma^2c\sim 3\,r_{12}(\Gamma/100)^2$~ms requires a small emission radius $r\simlt 10^{12}$~cm.  In fact, the self-similar tail material cannot be present at such small radii. The picture in \cite{Metzger19} leaves out the presence of the continual spindown wind between the flares. As explained in \Sect~\ref{bubble}, the wind applies pressure $P\sim \Lw/4\pi r^2c$, sweeps the slow tail out from the inner region, and inflates a hot wind bubble behind it (Figure~\ref{fig:bubble}). Furthermore, before a blast wave from the second flare could reach the cold tail of the first flare, it must strongly decelerate in the wind bubble (Figure~\ref{fig:bw_bubble}), where the shock maser mechanism is disabled by the relativistic temperature of the bubble.
\medskip

(3) There is another issue for the model of FRB emission by the blast wave in a slowly expanding medium: induced downscattering (\Sect~\ref{ind}). We find that it does not affect the maser emission only if the blast wave propagates in an ultra-relativistic wind, $\eta\simgt 50$.
\medskip

The discussion of induced scattering in \cite{Metzger19} is based on the estimates of \cite{Lyubarsky08}. We note here the following. First, the results of \cite{Lyubarsky08} may not apply to radii $r\ll 10^{14}$~cm, because $a\gg 1$ at these radii (see \Eq~(\ref{eq:aFRB})); the induced scattering calculations are safe for waves with strength parameter $a<1$. Second, \cite{Lyubarsky08} estimated induced scattering {\it outside} the radio beam. It has a damping effect at frequencies $\nu$ where the spectral slope is harder than $+1$ ($g^\prime>0$). The observed FRB spectra normally have much softer slopes; and in this case the role of induced scattering outside the beam is less certain. A useful constraint is provided by scattering {\it inside} the beam (\Sect~\ref{ind}), which disfavors shocks in slowly expanding media as FRB sources. 
\\

For these reasons the present paper focused on blast waves propagating in the freely expanding spindown wind rather than the slow tail of a previous flare. The wind carries cold $e^\pm$ plasma and expands with ultra-relativistic $\Gw$. Too frequent pollution of the magnetar wind with slow ion ejecta (more frequently than $\sim 1$ per day) would make the FRB model problematic. It appears that not every flare efficiently ejects ions in the frequent repeater FRB~121102, indicating a significant energy threshold for massive ion ejection from magnetars.

\acknowledgements
I thank Brian Metzger, Yuri Levin, Yuri Lyubarsky, and Lorenzo Sironi for comments on the manuscript.
This work is supported by NASA grant NNX17AK37G, Simons Foundation grant \#446228, and the Humboldt Foundation.

 \bibliography{frb}

\begin{thebibliography}{}
\expandafter\ifx\csname natexlab\endcsname\relax\def\natexlab#1{#1}\fi

\bibitem[{{Amato} \& {Arons}(2006)}]{Amato06}
{Amato}, E., \& {Arons}, J. 2006, \apj, 653, 325

\bibitem[{{Bannister} {et~al.}(2019){Bannister}, {Deller}, {Phillips},
  {Macquart}, {Prochaska}, {Tejos}, {Ryder}, {Sadler}, {Shannon}, {Simha},
  {Day}, {McQuinn}, {North-Hickey}, {Bhandari}, {Arcus}, {Bennert}, {Burchett},
  {Bouwhuis}, {Dodson}, {Ekers}, {Farah}, {Flynn}, {James}, {Kerr}, {Lenc},
  {Mahony}, {O{\textquoteright}Meara}, {Os{\l}owski}, {Qiu}, {Treu}, {U},
  {Bateman}, {Bock}, {Bolton}, {Brown}, {Bunton}, {Chippendale}, {Cooray},
  {Cornwell}, {Gupta}, {Hayman}, {Kesteven}, {Koribalski}, {MacLeod},
  {McClure-Griffiths}, {Neuhold}, {Norris}, {Pilawa}, {Qiao}, {Reynolds},
  {Roxby}, {Shimwell}, {Voronkov}, \& {Wilson}}]{Bannister19}
{Bannister}, K.~W., {Deller}, A.~T., {Phillips}, C., {et~al.} 2019, Science,
  365, 565

\bibitem[{{Beloborodov}(2009)}]{Beloborodov09}
{Beloborodov}, A.~M. 2009, \apj, 703, 1044

\bibitem[{{Beloborodov}(2013{\natexlab{a}})}]{Beloborodov13b}
---. 2013{\natexlab{a}}, \apj, 777, 114

\bibitem[{{Beloborodov}(2013{\natexlab{b}})}]{Beloborodov13a}
---. 2013{\natexlab{b}}, \apj, 762, 13

\bibitem[{{Beloborodov}(2017{\natexlab{a}})}]{Beloborodov17b}
---. 2017{\natexlab{a}}, \apjl, 843, L26

\bibitem[{{Beloborodov}(2017{\natexlab{b}})}]{Beloborodov17a}
---. 2017{\natexlab{b}}, \apj, 838, 125

\bibitem[{{Beloborodov} \& {Li}(2016)}]{Beloborodov16}
{Beloborodov}, A.~M., \& {Li}, X. 2016, \apj, 833, 261

\bibitem[{{Beloborodov} \& {Thompson}(2007)}]{Beloborodov07}
{Beloborodov}, A.~M., \& {Thompson}, C. 2007, \apj, 657, 967

\bibitem[{{Blandford} \& {Scharlemann}(1975)}]{Blandford75}
{Blandford}, R.~D., \& {Scharlemann}, E.~T. 1975, \apss, 36, 303

\bibitem[{{Bogovalov}(1999)}]{Bogovalov99}
{Bogovalov}, S.~V. 1999, \aap, 349, 1017

\bibitem[{{Buckley}(1977)}]{Buckley77}
{Buckley}, R. 1977, \mnras, 180, 125

\bibitem[{{Carrasco} {et~al.}(2019){Carrasco}, {Vigan{\`o}}, {Palenzuela}, \&
  {Pons}}]{Carrasco19}
{Carrasco}, F., {Vigan{\`o}}, D., {Palenzuela}, C., \& {Pons}, J.~A. 2019,
  \mnras, 484, L124

\bibitem[{{Cerutti} \& {Beloborodov}(2017)}]{Cerutti17}
{Cerutti}, B., \& {Beloborodov}, A.~M. 2017, \ssr, 207, 111

\bibitem[{{Cerutti} \& {Philippov}(2017)}]{Cerutti17b}
{Cerutti}, B., \& {Philippov}, A.~A. 2017, \aap, 607, A134

\bibitem[{{Chen} \& {Beloborodov}(2014)}]{Chen14}
{Chen}, A.~Y., \& {Beloborodov}, A.~M. 2014, \apjl, 795, L22

\bibitem[{{Connor}(2019)}]{Connor19}
{Connor}, L. 2019, \mnras, 487, 5753

\bibitem[{{Coppi} {et~al.}(1993){Coppi}, {Blandford}, \& {Rees}}]{Coppi93}
{Coppi}, P., {Blandford}, R.~D., \& {Rees}, M.~J. 1993, \mnras, 262, 603

\bibitem[{{Cordes} {et~al.}(2017){Cordes}, {Wasserman}, {Hessels}, {Lazio},
  {Chatterjee}, \& {Wharton}}]{Cordes17}
{Cordes}, J.~M., {Wasserman}, I., {Hessels}, J.~W.~T., {et~al.} 2017, \apj,
  842, 35

\bibitem[{{Coroniti}(1990)}]{Coroniti90}
{Coroniti}, F.~V. 1990, \apj, 349, 538

\bibitem[{{Drenkhahn} \& {Spruit}(2002)}]{Drenkhahn02}
{Drenkhahn}, G., \& {Spruit}, H.~C. 2002, \aap, 391, 1141

\bibitem[{{Duncan} \& {Thompson}(1992)}]{Duncan92}
{Duncan}, R.~C., \& {Thompson}, C. 1992, \apjl, 392, L9

\bibitem[{{Gaensler} {et~al.}(2005){Gaensler}, {Kouveliotou}, {Gelfand},
  {Taylor}, {Eichler}, {Wijers}, {Granot}, {Ramirez-Ruiz}, {Lyubarsky},
  {Hunstead}, {Campbell-Wilson}, {van der Horst}, {McLaughlin}, {Fender},
  {Garrett}, {Newton-McGee}, {Palmer}, {Gehrels}, \& {Woods}}]{Gaensler05}
{Gaensler}, B.~M., {Kouveliotou}, C., {Gelfand}, J.~D., {et~al.} 2005, \nat,
  434, 1104

\bibitem[{{Gallant} {et~al.}(1992){Gallant}, {Hoshino}, {Langdon}, {Arons}, \&
  {Max}}]{Gallant92}
{Gallant}, Y.~A., {Hoshino}, M., {Langdon}, A.~B., {Arons}, J., \& {Max}, C.~E.
  1992, \apj, 391, 73

\bibitem[{{Gelfand} {et~al.}(2005){Gelfand}, {Lyubarsky}, {Eichler},
  {Gaensler}, {Taylor}, {Granot}, {Newton-McGee}, {Ramirez-Ruiz},
  {Kouveliotou}, \& {Wijers}}]{Gelfand05}
{Gelfand}, J.~D., {Lyubarsky}, Y.~E., {Eichler}, D., {et~al.} 2005, \apjl, 634,
  L89

\bibitem[{{Goldreich} \& {Julian}(1969)}]{Goldreich69}
{Goldreich}, P., \& {Julian}, W.~H. 1969, \apj, 157, 869

\bibitem[{{Goldreich} \& {Julian}(1970)}]{Goldreich70}
---. 1970, \apj, 160, 971

\bibitem[{{Goldreich} \& {Reisenegger}(1992)}]{Goldreich92}
{Goldreich}, P., \& {Reisenegger}, A. 1992, \apj, 395, 250

\bibitem[{{Granot} {et~al.}(2011){Granot}, {Komissarov}, \&
  {Spitkovsky}}]{Granot11}
{Granot}, J., {Komissarov}, S.~S., \& {Spitkovsky}, A. 2011, \mnras, 411, 1323

\bibitem[{{Granot} {et~al.}(2006){Granot}, {Ramirez-Ruiz}, {Taylor}, {Eichler},
  {Lyubarsky}, {Wijers}, {Gaensler}, {Gelfand}, \& {Kouveliotou}}]{Granot06}
{Granot}, J., {Ramirez-Ruiz}, E., {Taylor}, G.~B., {et~al.} 2006, \apj, 638,
  391

\bibitem[{{Gruzinov} \& {Levin}(2019)}]{Gruzinov19}
{Gruzinov}, A., \& {Levin}, Y. 2019, \apj, 876, 74

\bibitem[{{Hakobyan} {et~al.}(2019){Hakobyan}, {Philippov}, \&
  {Spitkovsky}}]{Hakobyan19}
{Hakobyan}, H., {Philippov}, A., \& {Spitkovsky}, A. 2019, \apj, 877, 53

\bibitem[{{Hessels} {et~al.}(2019){Hessels}, {Spitler}, {Seymour}, {Cordes},
  {Michilli}, {Lynch}, {Gourdji}, {Archibald}, {Bassa}, {Bower}, {Chatterjee},
  {Connor}, {Crawford}, {Deneva}, {Gajjar}, {Kaspi}, {Keimpema}, {Law},
  {Marcote}, {McLaughlin}, {Paragi}, {Petroff}, {Ransom}, {Scholz}, {Stappers},
  \& {Tendulkar}}]{Hessels19}
{Hessels}, J.~W.~T., {Spitler}, L.~G., {Seymour}, A.~D., {et~al.} 2019, \apjl,
  876, L23

\bibitem[{{Hoshino} \& {Arons}(1991)}]{Hoshino91}
{Hoshino}, M., \& {Arons}, J. 1991, Physics of Fluids B, 3, 818

\bibitem[{{Iwamoto} {et~al.}(2017){Iwamoto}, {Amano}, {Hoshino}, \&
  {Matsumoto}}]{Iwamoto17}
{Iwamoto}, M., {Amano}, T., {Hoshino}, M., \& {Matsumoto}, Y. 2017, \apj, 840,
  52

\bibitem[{{Iwamoto} {et~al.}(2018){Iwamoto}, {Amano}, {Hoshino}, \&
  {Matsumoto}}]{Iwamoto18}
---. 2018, \apj, 858, 93

\bibitem[{{Kaspi} \& {Beloborodov}(2017)}]{Kaspi17}
{Kaspi}, V.~M., \& {Beloborodov}, A.~M. 2017, \araa, 55, 261

\bibitem[{{Katz}(2016)}]{Katz16}
{Katz}, J.~I. 2016, \apj, 826, 226

\bibitem[{{Kirk} {et~al.}(2009){Kirk}, {Lyubarsky}, \& {Petri}}]{Kirk09}
{Kirk}, J.~G., {Lyubarsky}, Y., \& {Petri}, J. 2009, in Astrophysics and Space
  Science Library, Vol. 357, Astrophysics and Space Science Library, ed.
  W.~{Becker}, 421

\bibitem[{{Kumar} {et~al.}(2017){Kumar}, {Lu}, \& {Bhattacharya}}]{Kumar17}
{Kumar}, P., {Lu}, W., \& {Bhattacharya}, M. 2017, \mnras, 468, 2726

\bibitem[{{Langdon} {et~al.}(1988){Langdon}, {Arons}, \& {Max}}]{Langdon88}
{Langdon}, A.~B., {Arons}, J., \& {Max}, C.~E. 1988, \prl, 61, 779

\bibitem[{{Law} {et~al.}(2017){Law}, {Abruzzo}, {Bassa}, {Bower},
  {Burke-Spolaor}, {Butler}, {Cantwell}, {Carey}, {Chatterjee}, {Cordes},
  {Demorest}, {Dowell}, {Fender}, {Gourdji}, {Grainge}, {Hessels}, {Hickish},
  {Kaspi}, {Lazio}, {McLaughlin}, {Michilli}, {Mooley}, {Perrott}, {Ransom},
  {Razavi-Ghods}, {Rupen}, {Scaife}, {Scott}, {Scholz}, {Seymour}, {Spitler},
  {Stovall}, {Tendulkar}, {Titterington}, {Wharton}, \& {Williams}}]{Law17}
{Law}, C.~J., {Abruzzo}, M.~W., {Bassa}, C.~G., {et~al.} 2017, \apj, 850, 76

\bibitem[{{Levinson}(2010)}]{Levinson10}
{Levinson}, A. 2010, \apj, 720, 1490

\bibitem[{{Lyubarsky}(2008)}]{Lyubarsky08}
{Lyubarsky}, Y. 2008, \apj, 682, 1443

\bibitem[{{Lyubarsky}(2014)}]{Lyubarsky14}
---. 2014, \mnras, 442, L9

\bibitem[{{Lyubarsky} \& {Eichler}(2001)}]{Lyubarsky2001b}
{Lyubarsky}, Y., \& {Eichler}, D. 2001, \apj, 562, 494

\bibitem[{{Lyubarsky} \& {Kirk}(2001)}]{Lyubarsky01}
{Lyubarsky}, Y., \& {Kirk}, J.~G. 2001, \apj, 547, 437

\bibitem[{{Lyutikov}(2010)}]{Lyutikov10}
{Lyutikov}, M. 2010, \pre, 82, 056305

\bibitem[{{Macquart} {et~al.}(2019){Macquart}, {Shannon}, {Bannister}, {James},
  {Ekers}, \& {Bunton}}]{Macquart19}
{Macquart}, J.~P., {Shannon}, R.~M., {Bannister}, K.~W., {et~al.} 2019, \apjl,
  872, L19

\bibitem[{{Margalit} \& {Metzger}(2018)}]{Margalit18}
{Margalit}, B., \& {Metzger}, B.~D. 2018, \apjl, 868, L4

\bibitem[{{Metzger} {et~al.}(2019){Metzger}, {Margalit}, \&
  {Sironi}}]{Metzger19}
{Metzger}, B.~D., {Margalit}, B., \& {Sironi}, L. 2019, \mnras, 485, 4091

\bibitem[{{Michel}(1969)}]{Michel69}
{Michel}, F.~C. 1969, \apj, 158, 727

\bibitem[{{Michel}(1971)}]{Michel71}
---. 1971, Comments on Astrophysics and Space Physics, 3, 80

\bibitem[{{Michel}(1973)}]{Michel73}
---. 1973, \apjl, 180, L133

\bibitem[{{Michilli} {et~al.}(2018){Michilli}, {Seymour}, {Hessels}, {Spitler},
  {Gajjar}, {Archibald}, {Bower}, {Chatterjee}, {Cordes}, {Gourdji}, {Heald},
  {Kaspi}, {Law}, {Sobey}, {Adams}, {Bassa}, {Bogdanov}, {Brinkman},
  {Demorest}, {Fernand ez}, {Hellbourg}, {Lazio}, {Lynch}, {Maddox}, {Marcote},
  {McLaughlin}, {Paragi}, {Ransom}, {Scholz}, {Siemion}, {Tendulkar}, {van
  Rooy}, {Wharton}, \& {Whitlow}}]{Michilli18}
{Michilli}, D., {Seymour}, A., {Hessels}, J.~W.~T., {et~al.} 2018, \nat, 553,
  182

\bibitem[{{Murase} {et~al.}(2016){Murase}, {Kashiyama}, \&
  {M{\'e}sz{\'a}ros}}]{Murase16}
{Murase}, K., {Kashiyama}, K., \& {M{\'e}sz{\'a}ros}, P. 2016, \mnras, 461,
  1498

\bibitem[{{Palmer} {et~al.}(2005){Palmer}, {Barthelmy}, {Gehrels}, {Kippen},
  {Cayton}, {Kouveliotou}, {Eichler}, {Wijers}, {Woods}, {Granot}, {Lyubarsky},
  {Ramirez-Ruiz}, {Barbier}, {Chester}, {Cummings}, {Fenimore}, {Finger},
  {Gaensler}, {Hullinger}, {Krimm}, {Markwardt}, {Nousek}, {Parsons}, {Patel},
  {Sakamoto}, {Sato}, {Suzuki}, \& {Tueller}}]{Palmer05}
{Palmer}, D.~M., {Barthelmy}, S., {Gehrels}, N., {et~al.} 2005, \nat, 434, 1107

\bibitem[{{Parfrey} {et~al.}(2013){Parfrey}, {Beloborodov}, \&
  {Hui}}]{Parfrey13}
{Parfrey}, K., {Beloborodov}, A.~M., \& {Hui}, L. 2013, \apj, 774, 92

\bibitem[{{Petroff} {et~al.}(2019){Petroff}, {Hessels}, \&
  {Lorimer}}]{Petroff19}
{Petroff}, E., {Hessels}, J.~W.~T., \& {Lorimer}, D.~R. 2019, \aapr, 27, 4

\bibitem[{{Philippov} {et~al.}(2015){Philippov}, {Spitkovsky}, \&
  {Cerutti}}]{Philippov15}
{Philippov}, A.~A., {Spitkovsky}, A., \& {Cerutti}, B. 2015, \apjl, 801, L19

\bibitem[{{Plotnikov} \& {Sironi}(2019)}]{Plotnikov19}
{Plotnikov}, I., \& {Sironi}, L. 2019, \mnras, 485, 3816

\bibitem[{{Popov} \& {Postnov}(2013)}]{Popov13}
{Popov}, S.~B., \& {Postnov}, K.~A. 2013, ArXiv e-prints, arXiv:1307.4924

\bibitem[{{Prochaska} {et~al.}(2019){Prochaska}, {Macquart}, {McQuinn},
  {Simha}, {Shannon}, {Day}, {Marnoch}, {Ryder}, {Deller}, {Bannister},
  {Bhandari}, {Bordoloi}, {Bunton}, {Cho}, {Flynn}, {Mahony}, {Phillips},
  {Qiu}, \& {Tejos}}]{Prochaska19}
{Prochaska}, J.~X., {Macquart}, J.-P., {McQuinn}, M., {et~al.} 2019, Science,
  366, 231

\bibitem[{{Ravi} {et~al.}(2019){Ravi}, {Catha}, {D'Addario}, {Djorgovski},
  {Hallinan}, {Hobbs}, {Kocz}, {Kulkarni}, {Shi}, {Vedantham}, {Weinreb}, \&
  {Woody}}]{Ravi19}
{Ravi}, V., {Catha}, M., {D'Addario}, L., {et~al.} 2019, \nat, 572, 352

\bibitem[{{Spruit}(2008)}]{Spruit08}
{Spruit}, H.~C. 2008, in American Institute of Physics Conference Series, Vol.
  983, 40 Years of Pulsars: Millisecond Pulsars, Magnetars and More, ed.
  C.~{Bassa}, Z.~{Wang}, A.~{Cumming}, \& V.~M. {Kaspi}, 391--398

\bibitem[{{Tendulkar} {et~al.}(2016){Tendulkar}, {Kaspi}, \&
  {Patel}}]{Tendulkar16}
{Tendulkar}, S.~P., {Kaspi}, V.~M., \& {Patel}, C. 2016, \apj, 827, 59

\bibitem[{{Tendulkar} {et~al.}(2017){Tendulkar}, {Bassa}, {Cordes}, {Bower},
  {Law}, {Chatterjee}, {Adams}, {Bogdanov}, {Burke-Spolaor}, {Butler},
  {Demorest}, {Hessels}, {Kaspi}, {Lazio}, {Maddox}, {Marcote}, {McLaughlin},
  {Paragi}, {Ransom}, {Scholz}, {Seymour}, {Spitler}, {van Langevelde}, \&
  {Wharton}}]{Tendulkar17}
{Tendulkar}, S.~P., {Bassa}, C.~G., {Cordes}, J.~M., {et~al.} 2017, \apjl, 834,
  L7

\bibitem[{{Thompson}(2008)}]{Thompson08b}
{Thompson}, C. 2008, \apj, 688, 499

\bibitem[{{Thompson}(2017)}]{Thompson17}
---. 2017, \apj, 844, 162

\bibitem[{{Thompson} \& {Duncan}(2001)}]{Thompson01}
{Thompson}, C., \& {Duncan}, R.~C. 2001, \apj, 561, 980

\bibitem[{{Vedantham} \& {Ravi}(2019)}]{Vedantham19}
{Vedantham}, H.~K., \& {Ravi}, V. 2019, \mnras, 485, L78

\bibitem[{{Wilson} \& {Rees}(1978)}]{Wilson78}
{Wilson}, D.~B., \& {Rees}, M.~J. 1978, \mnras, 185, 297

\bibitem[{{Zel'Dovich} \& {Levich}(1969)}]{Zeldovich69}
{Zel'Dovich}, Y.~B., \& {Levich}, E.~V. 1969, Soviet Journal of Experimental
  and Theoretical Physics, 28, 1287

\bibitem[{{Zhang}(2018)}]{Zhang18}
{Zhang}, B. 2018, \apjl, 854, L21

\bibitem[{{Zrake}(2016)}]{Zrake16}
{Zrake}, J. 2016, \apj, 823, 39

\end{thebibliography}

% \bibliography{ms.bbl}

\end{document}